\documentclass[apj, twocolumn, numberedappendix, twocolappendix]{openjournal}
\usepackage{newtxtext,newtxmath}
\usepackage[T1]{fontenc}
\usepackage{graphicx}	% Including figure files
\usepackage{amsmath}	% Advanced maths commands
\usepackage[breaklinks,colorlinks,citecolor=blue,urlcolor=blue]{hyperref}
\usepackage{orcidlink}
\usepackage{float,comment}

\usepackage{soul}

\DeclareRobustCommand{\VAN}[3]{#2}
\let\VANthebibliography\thebibliography
\def\thebibliography{\DeclareRobustCommand{\VAN}[3]{##3}\VANthebibliography}

\newcommand{\zspec}{z_{\rm spec}}
\newcommand{\zphot}{\hat z_{\rm photo}}
\newcommand{\SL}{{\it StratLearn} }
\newcommand{\SLz}{{\it StratLearn-z} }
\newcommand{\SLznospace}{{\it StratLearn-z}}

\begin{document}
\title[SL for photometric redshift]{\textit{StratLearn-z}: Improved photo-$z$ estimation from spectroscopic data subject to selection effects }

\author{\vspace{-1.0cm}Chiara Moretti$^{1,2,3,4}$\,\orcidlink{0000-0003-3314-8936}}
\email{cmoretti@sissa.it}
\author{Maximilian Autenrieth$^{5,6,7}$\,\orcidlink{0009-0006-2068-5950}}
\author{Riccardo Serra$^{1}$}
\author{Roberto Trotta$^{1,2,3,8}$}
\author{David A. van Dyk$^{5}$\,\orcidlink{0000-0002-0816-331X}}
\author{Andrei~Mesinger$^{2,9}$ }

\affiliation{$^{1}$SISSA -  International School for Advanced Studies, Via Bonomea 265, 34136 Trieste, Italy}
\affiliation{$^{2}$Centro Nazionale ``High Performance Computer, Big Data and Quantum Computing''}
\affiliation{$^{3}$INAF – Osservatorio Astronomico di Trieste, Via Tiepolo 11, I-34143 - Trieste, Italy}
\affiliation{$^{4}$INFN sezione di Trieste}
\affiliation{$^{5}$Statistics Section, Department of Mathematics, Imperial College London, 180 Queen’s Gate, London SW7 2AZ, UK}
\affiliation{$^{6}$Statistical Laboratory, DPMMS,
University of Cambridge, Wilberforce Road, Cambridge CB3 0WB, UK}
\affiliation{$^{7}$Institute of Astronomy and Kavli Institute for Cosmology, Madingley Road, Cambridge CB3 0HA, UK}
\affiliation{$^{8}$Department of Physics, Imperial College London, Blackett Laboratory, Prince Consort Rd, SW7 2AZ London, UK}
\affiliation{$^{9}$Scuola Normale Superiore (SNS), Piazza dei Cavalieri 7, Pisa, PI, 56125, Ialy}

%% % These dates will be filled out by the publisher
%% \date{Accepted XXX. Received YYY; in original form ZZZ}

%% % Enter the current year, for the copyright statements etc.
%% \pubyear{2023}

%% % Don't change these lines
%% \begin{document}
%% \label{firstpage}
%% \pagerange{\pageref{firstpage}--\pageref{lastpage}}

\begin{abstract}
A precise measurement of photometric redshifts (photo-$z$) is crucial for the success of modern photometric galaxy surveys. Machine learning (ML) methods show great promise in this context, but suffer from covariate shift in training sets due to selection bias where interesting sources, e.g., high redshift objects, are underrepresented, and the corresponding ML models exhibit poor generalisation properties. We present an application of the \SL method to the estimation of photo-$z$ (\SLznospace), validating against simulations where we enforce the presence of covariate shift to different degrees. \SL is a statistically principled approach which relies on splitting the combined source and target datasets into strata, based on estimated propensity scores. The latter is the probability for an object in the dataset to be in the source set, given its observed covariates.
After stratification, two conditional density estimators are fit separately within each stratum, and then combined via a weighted average. We benchmark our results against the GPz algorithm, quantifying the performance of the two algorithms with a set of metrics. Our results show that the \SLz metrics are only marginally affected by the presence of covariate shift, while GPz shows a significant degradation of performance, specifically concerning the photo-$z$ prediction for fainter objects for which there is little training data. In particular, for the strongest covariate shift scenario considered, \SLz yields a reduced fraction of catastrophic errors, a factor of 2 improvement for the RMSE as well as one order of magnitude improvement on the bias. We also assess the quality of the predicted conditional redshift estimates using the probability integral transform (PIT) and the continuous rank probability score (CRPS). The PIT for \SLz indicates that predictions are well-centered around the true redshift value, if conservative in their variance; the CRPS shows marked improvement at high redshifts when compared with GPz. Our {\sc julia} implementation of the method, \SLz, is publicly available at \url{https://github.com/chiaramoretti/StratLearn-z}.
\end{abstract}
\maketitle

%%%%%%%%%%%%%%%%% BODY OF PAPER %%%%%%%%%%%%%%%%%%

\section{Introduction}
The main science driver for current and planned cosmological experiments is the exploration of the dark sector. Stage-IV Dark Energy surveys, such as Euclid \citep{laureijs2011, euclidoverview2024}, the Vera C. Rubin Observatory Legacy Survey of Space and Time (LSST, \cite{lsst2018, ivezic2019}), and Roman Space Telescope \citep{akeson2019}, will pursue this goal by mapping the Universe over unprecedented volumes, delivering high precision measurements of cosmological observables.  Of key importance for such measurements is our ability to produce accurate estimates of redshift for billions of sources.

The most precise way to estimate redshifts is via spectroscopic observations. These, however, are demanding in terms of observational time, and challenging for faint, high-redshift objects: the sheer number of sources in future surveys will prevent spectroscopic follow-up for the vast majority. A viable and well established alternative is provided by the so-called photometric redshift (photo-$z$, see e.g., \cite{salvato2019, brescia2021, newman2022} for detailed reviews), which are extracted from broadband flux measurements performed at different wavelengths. While spectroscopy allows for highly accurate redshift measurements for individual objects, the sample obtained is often biased and unrepresentative of the full population, whereas photometric redshifts, despite being less precise for individual objects, allow the measurement of several objects simultaneously.

Accuracy and precision in the estimation of photo-$z$ are essential for the success of both current and future imaging surveys. In fact, systematic errors in the determination of redshifts from photometric observations can introduce biases into inferred cosmological parameters. Specifically, the lensing analysis relies on the construction of tomographic redshift bins, with minimal overlap between adjacent bins and an accurate determination of the mean redshift of each bin \citep{ma2006, amara2007}, as well as knowledge on the $n(z)$. Additionally, the clustering signal used in the 3x2pt analysis requires both the mean redshift and the width of the redshift distribution to be known with high accuracy \citep{tutusaus2020, porredon2022}.
There are two main approaches for estimating photo-$z$: template-fitting methods and machine learning (ML) based methods. The former are based on matching the observed photometry to catalogs of template galaxy spectra in order to extract the galaxy redshifts; these methods rely on the completeness of such catalogs. Publicly available template-fitting codes include {\sc LePhare} \citep{arnouts1999, ilbert2006}, {\sc BPZ} \citep{benitez2000}, {\sc Hyperz} \citep{bolzonella2000} and {\sc EAZY} \citep{brammer2008}. On the other hand, the advent of ML techniques has opened new ways to improve photo-$z$ estimation, taking advantage of artificial neural networks \citep{firth2003, collister2004, graff2014}  random forests \citep{carliles2010, carrascokind2013}, advanced ANNs \citep{sadeh2016}, Gaussian processes \citep{almosallam2016a, almosallam2016b}, nearest neighbors \citep{graham2018}, convolutional neural networks \citep{pasquet2019, henghes2022}, graph neural networks \citep{tosone2023}, and self-organizing maps \citep{carrascokind2014, masters2015}. 
Self-organizing maps project high-dimensional data points onto a two-dimensional grid.
During training, the grid nodes (or neurons) are iteratively adjusted to capture the underlying structure of the data. 
After training, photometric target galaxies are assigned to their closest grid nodes based on a distance metric (typically Euclidean distance), 
and photometric redshift (photo-$z$) estimates are derived by averaging the redshifts of the training galaxies associated with each node. This approach can be interpreted as a form of stratification of the data space, where each node effectively defines a stratum. 
In the original implementations by \cite{carrascokind2014} and \cite{masters2015}, the grids consist of approximately 1,000 to 11,000 nodes, making SOMz an extreme example of a stratified approach.

ML methods have successfully been applied to Stage-III surveys: both the Dark Energy Survey (DES, \cite{abbott2018}) and the Kilo-Degree Survey (KiDS, \cite{heymans2021}) have used self-organizing maps to estimate photo-$z$ (see \cite{myles2021} and \cite{hildebrandt2021} respectively), while the Hyper Suprime-Cam (HSC, \cite{aihara2018}) adopted conditional density estimators \citep{rau2023, sugiyama2023}. Performance analyses comparing different approaches have also been presented  \citep{hildebrandt2010, sanchez2014}. 
In preparation for Stage-IV surveys, several ML methods have been compared with each other and to template-fitting methods, in particular for Euclid \citep{desprez2020} and for LSST-DESC \citep{schmidt2020}. 

ML methods rely on source datasets used to train the algorithms where both spectroscopic and photometric observations are available, in order to learn the relationship between the observed photometry and redshift. A significant issue is  posed by non-representative training sets, i.e.,  spectroscopic training samples that are not random samples from the target photometric survey. Indeed, such non-representative samples are the usual situation in astronomy, where the lack of representativeness is a generic and widespread consequence of selection bias. An example is `Malmquist bias', \citep{malmquist1922, malmquist1925}), which results in spectroscopic catalogs being biased towards brighter, lower redshift sources. Observational biases can thus result in different distributions for the covariates of the source datasets, used to train the ML algorithms, and the target dataset, for which we wish to estimate photo-$z$. Under the assumption that the conditional distribution of the outcome (in this context, the redshift) given the covariates (colors/magnitudes) is the same in the source and target group, the effect is known as covariate shift \citep{moreno-torres2012} in the ML literature. The impact of covariate shift on photo-$z$ estimation has been investigated in \cite{freeman2017}, where the authors propose to mitigate the effect of covariate shift with the use of importance weights, and apply the proposed method to data from DR8 of the Sloan Digital Sky Survey. More recently, \cite{stylianou2022} investigates the performance of GPz \citep{almosallam2016a, almosallam2016b}, an algorithm based on Gaussian processes, applying it to simulations that feature imperfections in the spectroscopic training set. Additionally, \cite{sancipriano2023} studies the impact of incompleteness of the training set in the context of DES, while \cite{moskowitz2024} suggests a method to augment the training sample to improve photo-$z$ estimation.

In this work, we focus on \SL \citep{autenrieth2024a, autenrieth2024b}, a statistically motivated approach that improves the performance of ML-based algorithms in the presence of covariate shift. \SL splits the data into subgroups, or strata, based on the estimated propensity scores, which in the context of this paper is the probability of a galaxy in the data set being included in the spectroscopic source set, given its observed covariates (magnitudes or colors). Propensity scores are balancing scores, and a pivotal methodology in causal inference \citep{rosenbaum1983}. Conditioning on the propensity scores via stratification approximately balances within strata the covariates of source and target data, which then improves the fitting of ML methods within each stratum \citep{autenrieth2024a}.
In this paper we apply an implementation of the \SL method, \SLznospace, to photo-$z$ estimation, and assess its performance on simulated data with different degrees of covariate shift, with a view to future surveys such as Euclid and LSST.  Our goal is to showcase the ability of \SLz to provide accurate photo-$z$ estimates even in presence of strong covariate shift, which can reduce systematic biases and therefore be highly beneficial for such experiments.

Throughout, we refer to the implementation of \SL adopted here as \SLz.

%-------------------------------------------------------------------------------
\section{StratLearn -- Addressing Covariate Shift}
\label{sec:covariateshift}
In this section we present in more detail the covariate shift problem in the context of photo-$z$ estimation, and summarize the \SL approach developed in \cite{autenrieth2024a}.

\subsection{Covariate shift in non-representative spectroscopic data}
Starting from a source dataset of labelled observations with spectroscopic redshift $z^{(i)}$ and covariates $x^{(i)}$ (photometric magnitudes or colors), we aim to obtain redshift estimations for a much larger target set of unlabelled sources, i.e., for sources with only photometric information/covariates available. The source dataset $D_S = \{ (x_S^{(i)}, z_S^{(i)}) \}_{i=1}^{N_s}$ consists of $N_S$ galaxies sampled from the joint distribution $p_S(x,z)$, while the target dataset $D_T = \{ x_T^{(i)} \}_{i=1}^{N_T}$ consists of $N_T$ galaxies sampled from the joint distribution $p_T(x,z)$, with $N_T \gg N_S$. Selection effects induce differences between the source and target distributions, $p_S(x) \neq p_T(x)$, leading to $p_S(x,z) \neq p_T(x,z)$. However, we assume that the conditional distribution of redshifts given the covariates is the same for both source and target, $p_S(z|x) = p_T(z|x)$. This situation is known to affect the performances of ML algorithms, because the patterns learned from the (unrepresentative) source set do not generalise properly to the target set.
\footnote{We note that the assumption $p_S(z|x) = p_T(z|x)$ may not always hold, e.g. in case of additional selection bias in the identification of a reliable spectroscopic redshift \citep{hartley2020} (see also the discussion in Sec.~\ref{sec:conclusion}). We plan to explore such cases in a future work.}

\subsection{StratLearn}
\label{sec:stratlearn}

\SL is designed to improve the generalisation properties of ML algorithms in presence of covariate shift. While the method is general and applicable to both classification and regression problems, we focus here on photo-$z$ estimation, as discussed in the previous section. In the following we denote all estimated quantities with a hat symbol, e.g., $\hat p$.

The approach relies on stratification of data into sub-groups, or strata, based on propensity scores, i.e., the estimated probability for an object $i$ in the dataset to be in the source set ($s_i = 1$) given the observed covariates $x_i$:
\begin{equation}
  \label{eq:propensity-scores}
    e(x_i) = P(s_i=1 | x_i) \, ,
\end{equation}
with the assumption that $0<e(x_i)<1$.
For each object we obtain an estimate of the propensity score $\hat{e}(x_i)$ via binary classification of source and target data. Specifically, we employ a logistic regression model with all photometric magnitudes/colors as covariates, and group the data in $K$ strata based on the quantiles of the estimated propensity score distribution. We use $K=5$ strata, as it has been shown that this number is able to remove at least 90\% of the bias for many distributions \citep{cochran1968, rosenbaum1983}.  The stratification process balances the covariate distribution in the source and target sets \citep{rosenbaum1983}, resulting in $p_{T_j}(z,x) \simeq p_{S_j}(z,x)$, for $j = 1,\dots,K,$ where the subscript $T_j$ denotes conditioning on assignment to the $j$-th target stratum (and analogously, $S_j$ for source stratum).

We use a weighted average of two supervised full conditional density estimators, trained separately in each stratum, to obtain a non-parametric estimate of the full galaxy photo-$z$ conditional densities of each object in the target set, $\hat{p}_T(z\mid x)$, conditional on its observed covariates.  We adopt the {\it ker-NN} estimator proposed in \cite{freeman2017} and the {\it Series} estimator described in \cite{izbicki2016}: the former relies on a kernel smoothed histogram of the redshift of the {\it k} nearest neighbors to $x$, within each stratum, to compute the conditional redshift distribution, while the latter adapts a lower-dimensional subspace of the covariates, $x$, as the intrinsic dimension of the data, based on data-dependent eigenfunctions of a kernel based estimator.  Our choice of conditional density estimators is based on good performance shown in previous work, e.g., \cite{autenrieth2024a}. However, generally, the \SL framework could be combined with any model.

We fit the two models using the labelled spectroscopic source data as a training set, and obtain an estimate of the redshift probability distribution for each galaxy in the photometric target data. If the source set is representative of the target set, conditional density estimators commonly aim to minimize the generalized risk under the $L^2$-loss:
\begin{equation}
  \label{eq:loss}
  \hat{R}_S(\hat{p}) = \frac{1}{N_S}\sum_{k=1}^{N_S} \int \hat{p}^2(z\mid x_S^{(k)}) \mathrm{d}z - 2 \frac{1}{N_S}\sum_{k=1}^{N_S}\hat{p}(z_S^{(k)}\mid x_S^{(k)}) \, ,
\end{equation}
where the subscript $`S`$ indicates the source set, $z_S^{(k)}$ is the redshift and $x_S^{(k)}$ refers to the covariates for the $k$-th object in the source set.
We notice however that the presence of covariate shift requires minimisation of the {\it target} risk $\hat{R}_T(\hat{p})$ in order to obtain accurate target estimates, which in turn would require having access to the target redshifts. The stratification procedure described above allows for minimisation of $\hat{R}_{T_j}(\hat{p})$ via minimisation of $\hat{R}_{S_j}(\hat{p})$ within each stratum $j$ (see \cite{autenrieth2024b} for a more detailed description). Finally, the two conditional density estimators are combined via
\begin{equation}
  \label{eq:estimators-combination}
\hat{p}(z \mid x)  = (1-\gamma) \hat{p}_{\rm Series}(z\mid x) + \gamma~\hat{p}_{\rm ker-NN}(z\mid x) \, ,
\end{equation}
where $0 \leq \gamma \leq 1$ is a weight parameter. This requires a further optimisation to compute $\gamma$ (within each stratum) via minimisation of the generalised risk:
\begin{equation}
  \hat{R}_{S2_j}(\hat{p}) = \frac{1}{N_{T_j}} \sum_{k=1}^{N_{T_j}} \int \hat{p}^2(z \mid x_T^{(k)}) \mathrm{d}z - 2 \frac{1}{N_{S_j}} \sum_{k=1}^{N_{S_j}} \hat{p}(z_S^{(k)}\mid x_S^{(k)}) \, .  
\end{equation}

%-------------------------------------------------------------------------------
\section{Simulated Dataset}
\label{sec:data}

To assess the performance of \SL we make use of a simulated dataset derived from the Buzzard Flock\footnote{\url{https://buzzardflock.github.io/index.html}} simulation suite \citep{derose2019}, initially developed for DES. In particular, we adopt the LSST-DESC DC1 mock catalog used in \cite{stylianou2022}\footnote{Specifically, the catalogue distributed with the {\sc pzflow} package \citep{crenshaw2024} at \url{https://github.com/jfcrenshaw/pzflow/blob/main/pzflow/example_files/galaxy-data.pkl}}, consisting of a sample of 100,000 galaxies with redshift range $0<z<2.3$ and complete photometry in the {\it ugrizy} bands, from which we compute colors to serve as input covariates to our algorithm.

The dataset was constructed from the catalogs described in \cite{schmidt2020} (specifically, see Sec.~2 of that work and the references therein for a more detailed description of the dataset). \cite{stylianou2022} uses this dataset to study the impact of systematics in the training dataset on the GPz algorithm \citep{almosallam2016a, almosallam2016b}.  Particular attention is paid to two possible degradations in the training dataset that might affect the performance of the algorithm: redshift incompleteness and emission line confusion. The redshift incompleteness in particular is similar to what we explore here, we therefore use GPz to benchmark the performance of \SLznospace. However, we introduce incompleteness (and therefore, lack of representativeness) in the training data via a different prescription, as detailed in Sec.~\ref{sec:cs}.

\subsection{Introducing covariate shift}
\label{sec:cs}

We enforce covariate shift in the dataset by performing rejection sampling on the $r$-band, to obtain the target data, following \cite{izbicki2016}. This approach is designed to mimic a selection effect that biases the source set towards brighter objects, similar to what happens with real observations. Specifically, we resample the full dataset via rejection sampling, assigning each galaxy to the target set with probability $p(s = 0 \mid x) = f_{B(\alpha, \beta)}(x_r)/ \max_{x_r}  f_{B(\alpha, \beta)}(x_r)$, where $f_{B(\alpha, \beta)}$ is the density of a beta random variable with parameters $\alpha$, $\beta$ and the $r$-band magnitudes, $x_r$, are re-scaled to be between 0 and 1. Galaxies not assigned to the target set are assigned to the source data.

We consider four different scenarios with different degrees of covariate shift (CS):
\begin{itemize}
    \item no covariate shift: data are randomly split between source and target sets;
    \item weak covariate shift: $\alpha=4$, $\beta=5$;
    \item mild covariate shift: $\alpha=5$, $\beta=6$;
    \item strong covariate shift: $\alpha=5$, $\beta=7$.
\end{itemize}

\begin{figure}[t!]
\centering
\includegraphics[width=\columnwidth]{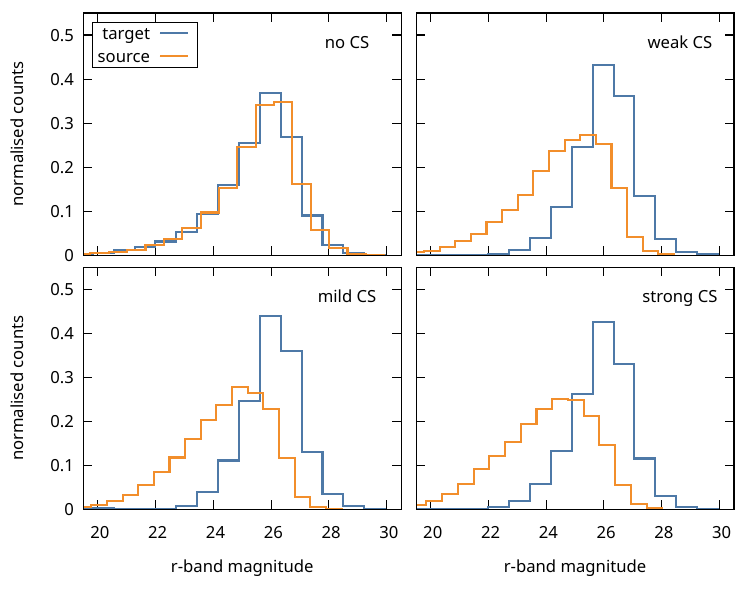}
\caption{Source and target normalised distributions of $r$-band magnitudes for the four covariate shift (CS) cases, as stated in each panel. Orange histograms represent the source (labelled) datasets, while in blue we plot the target (unlabelled) datasets. Notice how the rejection sampling procedure introduces a shift -- to different degrees, based on the chosen parameters (i.e. scenarios) -- between the source and target sets.}
\label{fig:test-train-histograms}
\end{figure}

\begin{figure}[t!]
\centering
\includegraphics[width=\columnwidth]{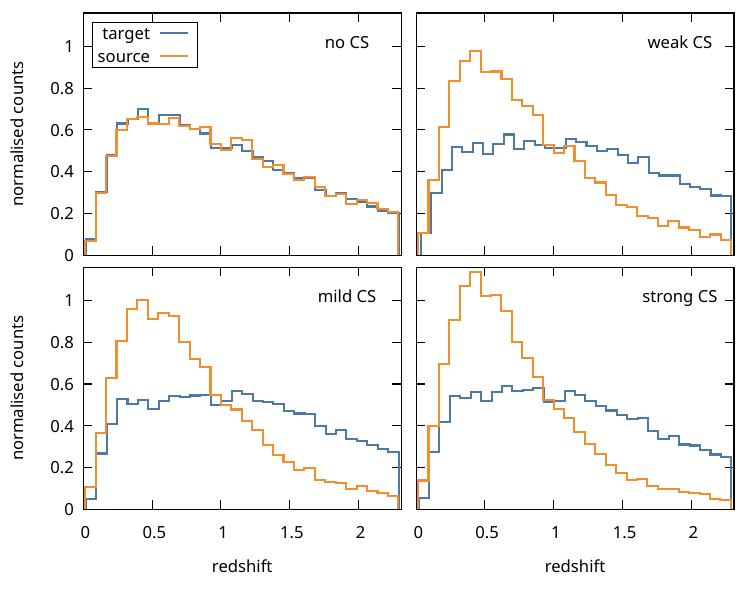}
\caption{Source and target normalised redshift distributions for the four covariate shift (CS) cases, as obtained after the rejection-sampling procedure described in Sec.~\ref{sec:cs}. Orange histograms represent the source datasets, while in blue we plot the target datasets. We plot the true redshift distribution for the target data, which in reality is of course unknown.}
\label{fig:test-train-redshift-hist}
\end{figure}

From this procedure we obtain four source datasets, three of which have $r$-band distributions that are shifted with respect to their corresponding target datasets, as shown in Fig.~\ref{fig:test-train-histograms} for the various covariate shift cases.  Although in a simplified form, the rejection sampling procedure mimics the effect of selection bias in real data, where the limiting magnitude of the $r$-band distribution of spectroscopic observations is usually at lower values than for photometric observations \citep[see e.g.,  Fig.~3 of][]{izbicki2016}, because of higher probability of detecting brighter objects.  Such bias is also visible as a shift in the redshift distributions for source and target, as shown in Fig.~\ref{fig:test-train-redshift-hist}: the source set features a peak at lower redshifts, that is more pronounced the stronger the covariate shift. We also plot the distributions for all other photometric bands in Appendix~\ref{app:band-distrib}.

The rejection sampling procedure described above results in source and target datasets of very different sizes among the four covariate shift scenarios. To focus on the impact of covariate shift, and remove dependence of the results on the size of the datasets, we want the source sets for all scenarios to have the same size (and similarly for the target sets). Thus, we randomly resample the sets to construct the source datasets, used to train the conditional density estimators, and target datasets, used to compute performance metrics. 

Specifically, we randomly sub-sample 20,000 objects for each training set and 63,000 in each target set, except the target dataset in the weak covariate shift scenario was not sub-sampled because rejection sampling yielded only 62,000 objects.

We choose the target sets to be significantly larger than the training sets to reproduce what happens with real data, where labels (spectroscopic redshifts) are only available for a small fraction of the total photometric measurements. For example, Euclid is expected to observe photometry for $\sim1.5$ billion galaxies, and will rely on complementary ground-based spectroscopic samples to calibrate the relation between galaxy colours and redshifts, with significant effort put into targeted spectroscopic observations to span the full galaxy colour space (see \cite{euclidoverview2024} for more details). 

\subsection{\SL computation}
\label{sec:sl-computation}

For each scenario, we estimate the propensity scores using a logistic regression model to predict the binary classification into the source and target data.  In particular, we use the $r$-band data and colors obtained from all other available photometric bands as covariates (i.e., $u-g$, $g-r$, $r-i$, $i-z$, $z-y$). 
We scale all covariates to have mean zero and standard deviation one.
We then partition each of the datasets (combining source and target data under each scenario) into five strata, based on the quintiles of the corresponding propensity score distributions. The resulting strata are populated as detailed in Table~\ref{tab:strata-balance}.  The no covariate shift case results in approximately equally sized strata: $\sim 4000$ in the source set and $\sim12500$ in the target set, per stratum.  For the datasets with covariate shift there is significantly more data in the first source strata (corresponding to a lower mean redshift), as expected.

\begin{table*}
\centering
\caption{Balance between strata after propensity score estimation, for the four covariate shift (CS) scenarios. We also report the mean redshift of each stratum for both the source and target datasets. A similar value for the mean redshift within strata indicates that the source and target data are balanced, although we note that this diagnostic is not possible in practice with real source/target data, as $z$ is not observed in the target dataset.}
\label{tab:strata-balance}
\resizebox{0.9\textwidth}{!}{%
\begin{tabular}{lrrrrrrrrrrr} 
\hline \hline
                & \multicolumn{2}{c}{No CS} && \multicolumn{2}{c}{weak CS} && \multicolumn{2}{c}{mild CS} && \multicolumn{2}{c}{strong CS}  \\ 
\cline{2-3} \cline{5-6} \cline{8-9} \cline{11-12}
 Stratum       & Size & Mean $z$ && Size & Mean $z$ && Size & Mean $z$ && Size & Mean $z$\\ 
\hline
1 (source)     & 4113  & 1.298  && 9959  & 0.625  && 10470 & 0.617  && 11038 & 0.571\\
1 (target)     & 12487 & 1.306  && 6441  & 0.708  && 6130  & 0.705  && 5562  & 0.666\\[3pt] %\hline 
2 (source)     & 4053  & 1.313  && 4479  & 0.899  && 4643  & 0.894  && 4764  & 0.850\\
2 (target)     & 12547 & 1.311  && 11921 & 0.904  && 11957 & 0.887  && 11836 & 0.854\\[3pt] %\hline 
3 (source)     & 3978  & 1.051  && 2799  & 0.979  && 2644  & 0.954  && 2432  & 0.963\\
3 (target)     & 12622 & 1.058  && 13601 & 0.984  && 13956 & 0.972  && 14168 & 0.962\\[3pt] %\hline 
4 (source)     & 3929  & 0.740  && 1917  & 1.113  && 1549  & 1.130  && 1253  & 1.061\\
4 (target)     & 12671 & 0.753  && 14483 & 1.121  && 15051 & 1.108  && 15347 & 1.075\\[3pt] %\hline 
5 (source)     & 3927  & 0.644  && 846   & 1.498  && 694   & 1.469  && 513   & 1.406\\
5 (target)     & 12673 & 0.624  && 15554 & 1.579  && 15906 & 1.585  && 16087 & 1.525\\[3pt] %\hline  
total (source) & 20000 & 1.014  && 20000 & 0.820  && 20000 & 0.795  && 20000 & 0.737\\
total (target) & 63000 & 1.009  && 62000 & 1.121  && 63000 & 1.117  && 63000 & 1.087\\
\hline\hline
\end{tabular}
}
\end{table*}

\begin{figure}[!t]
\centering
\includegraphics[width=0.95\columnwidth]{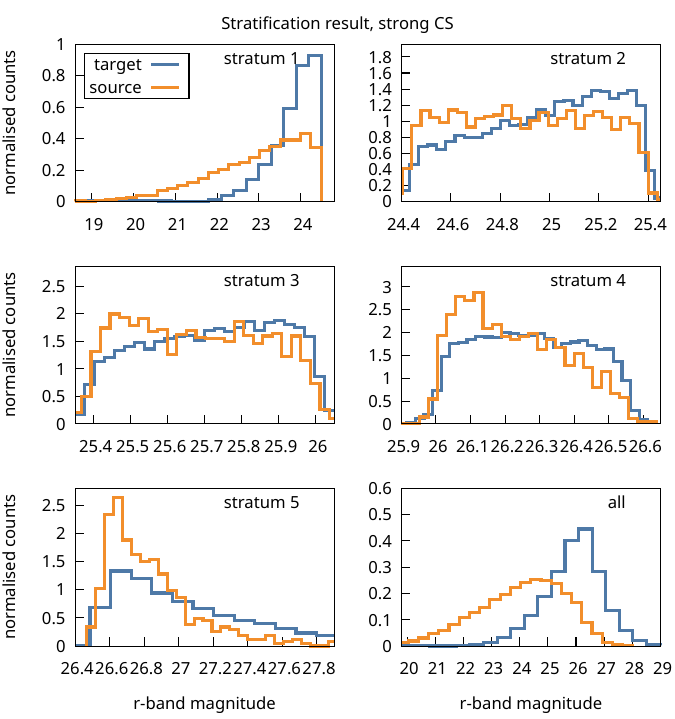}
\caption{Distribution of r-band magnitudes for source (orange histograms) and target (blue histograms) sets for the strong covariate shift (CS) scenario, after performing stratification based on estimated propensity scores. In the lower right panel, where we plot the full source and target sets, the distributions are significantly different, while in the other panels, representing each a different stratum, we are able to achieve improved balance.}
  \label{fig:rband-strata}
\end{figure}

Fig.~\ref{fig:rband-strata} plots the distributions of the $r$-band magnitudes for the source and target sets within each stratum of the strong covariate shift scenario.  Stratification achieves approximately balanced distributions within each strata. Additionally, stratification results in a redshift partition, where in each stratum the mean redshift of the source data approximately matches the mean redshift of the target data -- an indication that stratification is approximately removing covariate shift~\citep{autenrieth2024a} within each stratum. Redshift distribution for source and target sets for the strong covariate shift case are shown in Fig.~\ref{fig:redshift-strata} of Appendix \ref{app:stratification-results}, showcasing once again the effectiveness of \SL to improve balance within each stratum.

We train the ker-NN and {\it Series} conditional density estimators separately within each stratum, and combine them using Eq.~\ref{eq:estimators-combination} to obtain an estimate of the conditional distribution of redshift for each object in the target set. We emphasize that only the source redshifts are used to train the models, while the target redshifts are used exclusively to evaluate performance metrics.  The estimated (photometric) redshift $\zphot$, computed as the mean of the estimated conditional redshift distribution, is compared to the true (spectroscopic) redshift $\zspec$ for each galaxy in the target dataset, and performance in each covariate shift scenario is assessed using several metrics (see Sec.~\ref{sec:results}). As discussed in Sec.~\ref{sec:data}, we benchmark our results against the GPz code run on the same datasets. 

%-------------------------------------------------------------------------------
\section{\SLz performance}
\label{sec:results}

We evaluate the performance of \SLz in terms of both its point estimate $\zphot$ and its estimate of the conditional distribution of redshift via several metrics applied to the conditional densities obtained for the target dataset.

To obtain more robust numerical results for the performance metrics, we run 15 independent realisations for each covariate shift scenario. For each realisation, the datasets are constructed as follows:
\begin{itemize} \item rejection sampling, as described in Sec.~\ref{sec:cs}, is applied to the full dataset to generate four distinct source/target pairs corresponding to the different covariate shift scenarios; \item from each of these sets, we randomly sample 20,000 objects for the source set and 63,000 for the target set (62,000 for the weak covariate shift scenario; see Sec.~\ref{sec:cs}). \end{itemize}

The procedure is repeated 15 times, in order to obtain a total of 60 source/target sets, 15 for each covariate shift scenario.  The models are trained separately on each source set, and conditional density estimations constructed for the respective target set.  For each realisation, several performance metrics are then computed as described in the next section.

\subsection{Point estimate metrics}
\label{sec:point-metrics}

The quality of the photo-$z$ point estimate $\zphot$ is assessed by comparing it to the true redshift $\zspec$. Because we are working with simulated data, $\zspec$ is available for all objects in the target set, which can therefore be used to compute performance metrics.  We focus on the root mean square error (RMSE), the bias, and the fraction retained (FR). The latter provides an estimate of (100 times) the fraction of objects for which $\zphot$ does not exhibit catastrophic errors, defined as those objects for which residual differences between $\zphot$ and $\zspec$ are above some large threshold, defined relative to specific survey requirements. FR corresponds therefore to the fraction of redshift estimates that are not catastrophic, i.e., are `good' (see Eq.s~\ref{eq:fr15} and \ref{eq:fr05} for the thresholds adopted here). The metrics are evaluated as follows:

\begin{itemize}
\item Root mean square error (RMSE):
  \begin{equation}
    \sqrt{\frac{1}{N_T} \sum_{i=1}^{N_T} \left( \zspec^{(i)} - \hat{z}_{\rm photo}^{(i)} \right)^2} \, ;
    \label{eq:rmse}
  \end{equation}
  
\item mean error, commonly referred to in the literature as `bias':
  \begin{equation}
    \frac{1}{N_T} \sum_{i=1}^{N_T} \left( \zspec^{(i)} - \hat{z}_{\rm photo}^{(i)} \right) \, ;
    \label{eq:bias}
  \end{equation}

\item FR15: 
  \begin{equation}
    \frac{100}{N_T} \sum_{i=1}^{N_T} \left[ \, \left| \frac{\zspec^{(i)} - \hat{z}_{\rm photo}^{(i)}}{1+\zspec^{(i)}} \right| < 0.15 \right] \, ; \ \hbox{ and}
    \label{eq:fr15}
  \end{equation} 

\item FR05:
  \begin{equation}
    \frac{100}{N_T} \sum_{i=1}^{N_T} \left[ \, \left| \frac{\zspec^{(i)} - \hat{z}_{\rm photo}^{(i)}}{1+\zspec^{(i)}} \right| < 0.05 \right] \, .
    \label{eq:fr05}
  \end{equation}
\end{itemize}

The RMSE and mean error used in the context of photo-$z$ estimation are sometimes normalised by a factor $1+\zspec$: this serves to down-weigh the deviation from the spectroscopic redshift for high-redshift sources. For example, \cite{almosallam2016a} and \cite{desprez2020} use the normalised definitions, while \cite{hartley2020} and \cite{sancipriano2023} use the un-normalised ones. We note that the choice of performance metrics in the literature is far from homogeneous.  Given our benchmark against GPz, we adopt the metrics used in \cite{stylianou2022}, but report our results in terms of both the normalised and un-normalised metrics. 

\subsection{Conditional redshift distribution metrics}
\label{sec:pdf-metrics}

Measuring the precision of the conditional redshift distribution is more complicated, because there are no `true' conditional redshift distributions available. One possibility is using the distribution of the probability integral transform (PIT, \cite{dawid1984}, see also \cite{polsterer2016}), defined as
\begin{equation}
    {\rm PIT} = \int_0^{\zspec} \hat{p}_T(z|x) \mathrm{d}z \, ,
\end{equation}
where  $\hat{p}_T(z|x)$ is the conditional redshift distribution prediction for the target set given the observed photometry $x$. Under the assumption that $\hat{p}_T(z|x)$ exactly represents the conditional distribution of the true redshift, $\zspec$, given $x$, PIT is uniformly distributed over the unit interval.  On the other hand, shifts from uniformity in the PIT distribution suggest systematic deviations in the estimates of the conditional distributions of redshift, while over-estimation (under-estimation) of the spread of the conditional distribution of redshift, i.e., the uncertainty associated with the estimated $\zphot$, appear as a bump (dip) in the histogram of PIT. The PIT distribution is commonly used to visually evaluate the quality of estimated photo-$z$ distributions \citep[see e.g.][]{bordoloi2010, polsterer2016, disanto2018, tanaka2018, desprez2020}.

However, we note that the PIT distribution is not an indicator of the information content of the estimated conditional density. Indeed, \cite{schmidt2020} shows that a method with no predictive power can produce a perfectly uniform PIT distribution. As an example, the {\sc TrainZ} estimator described in \cite{schmidt2020} assigns the marginal redshift distribution of the source data to each galaxy in the target set. This distribution does not depend on the photometric data and produces the same photo-$z$ estimate for every galaxy in the target set, however, {\sc TrainZ} is able to achieve a perfect PIT.

Additionally, we compute the continuous rank probability score (CRPS, \cite{hersbach2000, polsterer2016}), defined as the integral of the squared difference between the redshift cumulative distribution function and the Heaviside step function centered at the true redshift value. More precisely, for each galaxy $i$, the CRPS is given by
\begin{equation}
    {\rm CRPS}^{(i)} = \int_{-\infty}^{+\infty} \left[ {\rm CDF}^{(i)}(z) - {\rm H}(z-\zspec^{(i)}) \right]^2 \mathrm{d}z, 
\end{equation}
where ${\rm CDF}^{(i)}(z)$ is the cumulative distribution function of the estimated conditional redshift distribution, 
and ${\rm H}(z - \zspec^{(i)})$ is the Heaviside step function,  i.e.,
\begin{equation}
    {\rm H}(z - \zspec^{(i)}) = 
    \begin{cases}
        0 & \text{if } z < \zspec^{(i)}, \\
        1 & \text{if } z \geq \zspec^{(i)}.
    \end{cases}
\end{equation}
The CRPS quantifies the discrepancy between the estimated conditional redshift distribution and a Dirac delta function at $\zspec^{(i)}$; it deviates from zero when the conditional distribution is broad or miscentered from the true redshift $\zspec^{(i)}$. The median CRPS across all objects in the target dataset serves as a measure of the overall quality of the estimated redshift distributions.

\subsection{Results and comparison to GPz}

We compare the performance of \SLz to the output of GPz \citep{almosallam2016a, almosallam2016b} for each of the four covariate shift scenarios, by running both codes on the same datasets\footnote{We use the GPz version distributed with \cite{stylianou2022} and available from \url{https://github.com/nataliastylianou/photo-$z$/tree/main}.}. GPz is based on sparse Gaussian processes and produces input-dependent (in this context, photometry-dependent) variance estimates which take into account both the density of the source data and noise in the photometric observations. One drawback of GPz is that the predicted conditional redshift distribution is forced to be a unimodal Gaussian.  The GPz code was compared to other existing photo-$z$ software in \cite{schmidt2020} for LSST and in the photometric challenge paper for Euclid \citep{desprez2020}; as is the case for other ML approaches, its performance is strongly dependent on the quality of the spectroscopic set used for training.

\begin{figure*}
  \includegraphics[width=0.5\textwidth]{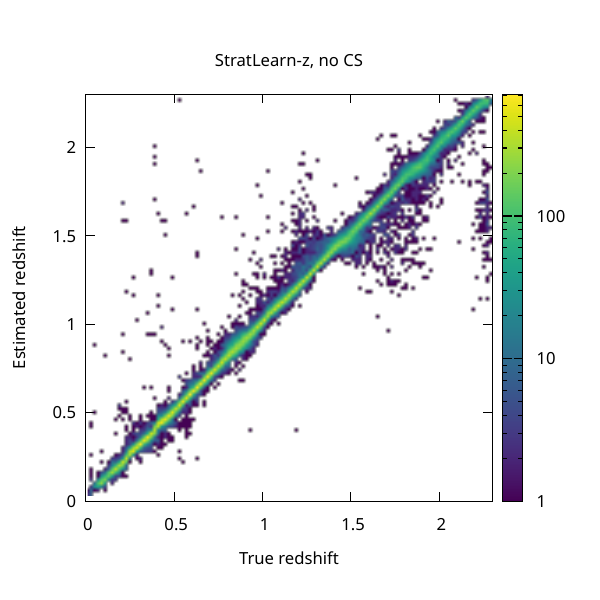} 
  \includegraphics[width=0.5\textwidth]{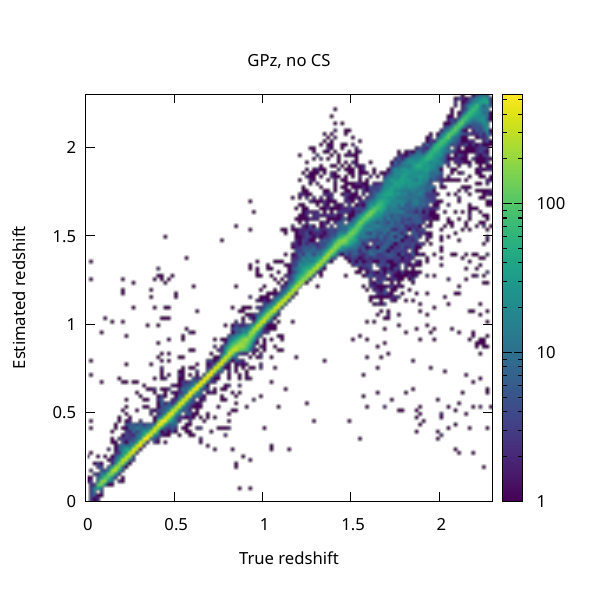} 
  \caption{2D histograms for the estimated $\zphot$ versus the true $\zspec$ for the no covariate shift (CS) case, $\alpha=1$, $\beta=1$. Left: \SLznospace, right: GPz. The colorbar indicates the number of sources in each 2D bin.}
  \label{fig:sl-result-no-cs}
\end{figure*}

\begin{figure*}
  \includegraphics[width=0.5\textwidth]{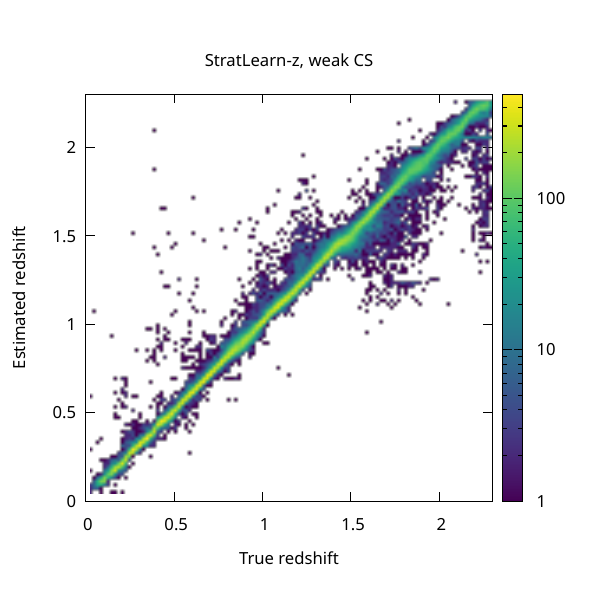} 
  \includegraphics[width=0.5\textwidth]{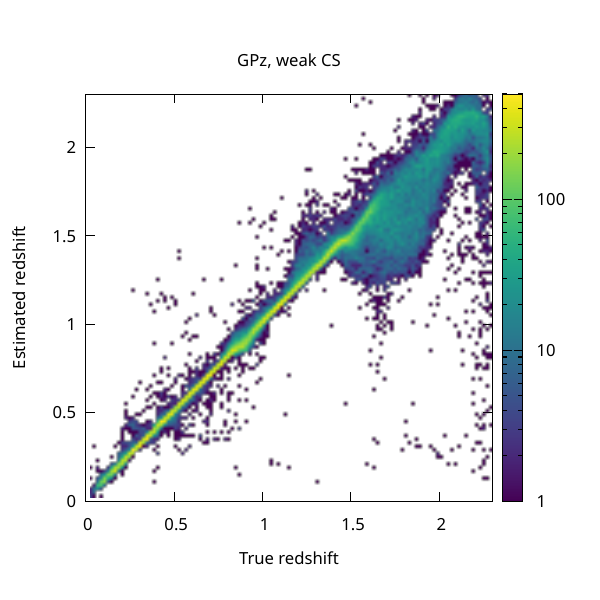} 
  \caption{Same as Fig.~\ref{fig:sl-result-no-cs}, but for the weak covariate shift case, $\alpha=4$, $\beta=5$. Left: \SLznospace, right: GPz. }
  \label{fig:sl-result-mild-cs-low}
\end{figure*}

\begin{figure*}
\includegraphics[width=0.5\textwidth]{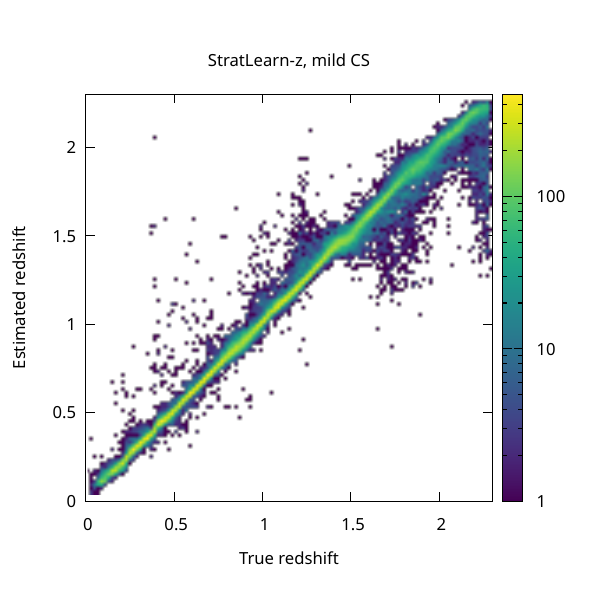} 
\includegraphics[width=0.5\textwidth]{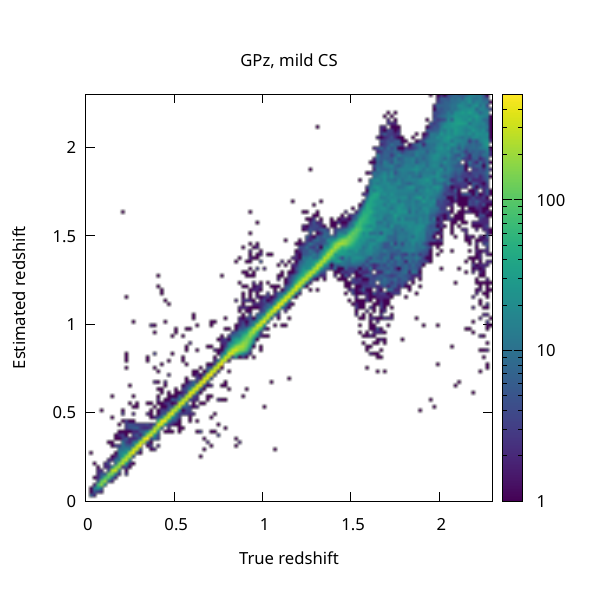} 
\caption{Same as Fig.~\ref{fig:sl-result-no-cs}, but for the mild covariate shift case, $\alpha=5$, $\beta=6$. Left: \SLznospace, right: GPz.}
\label{fig:sl-result-mild-cs-high}
\end{figure*}

\begin{figure*}
\includegraphics[width=0.5\textwidth]{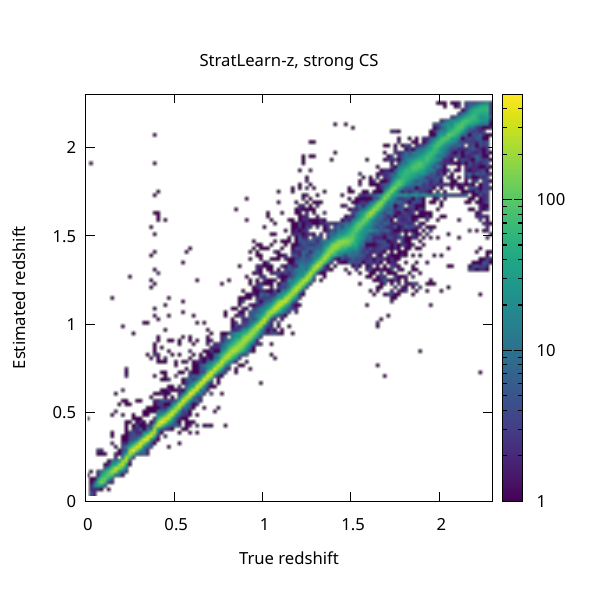}
\includegraphics[width=0.5\textwidth]{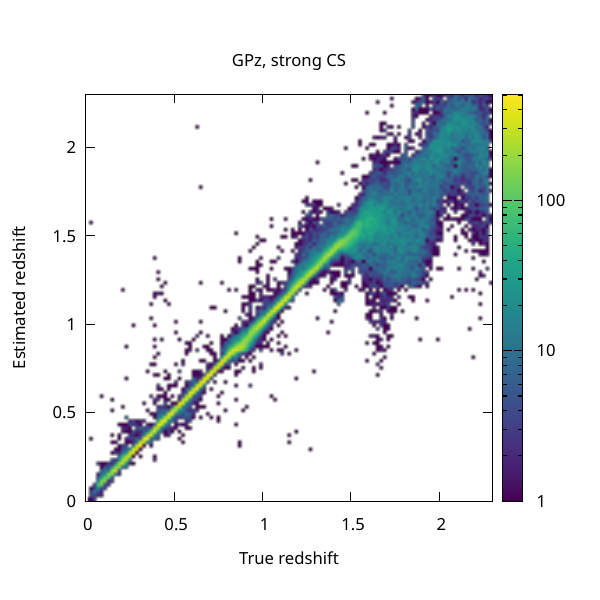}
\caption{Same as Fig.~\ref{fig:sl-result-no-cs}, but for the strong covariate shift case, $\alpha=5$, $\beta=7$. Left: \SLznospace, right: GPz.}
\label{fig:sl-result-strong-cs}
\end{figure*}

The 2D histograms in Fig.~\ref{fig:sl-result-no-cs}, \ref{fig:sl-result-mild-cs-low}, \ref{fig:sl-result-mild-cs-high}, and \ref{fig:sl-result-strong-cs} show the true $\zspec$ versus the estimated $\zphot$, obtained as the mean of the predicted conditional redshift distribution, for each covariate shift scenario and for both \SLz (left panels) and GPz (right panels). Visual inspection of the 2D histograms indicates that the GPz predictions are strongly degraded for high-redshift objects in the presence of covariate shift, where the distribution of $\zspec$-$\zphot$ exhibits high variance.  On the other hand, the \SLz predictions are only marginally impacted by covariate shift, even in the worst case scenario shown in Fig.~\ref{fig:sl-result-strong-cs}. We note that the plots refer to one specific realisation out of the 15 runs we perform for each case, however, a similar pattern is apparent for the remaining realisations.

\begin{figure*}
\includegraphics[width=0.5\textwidth]{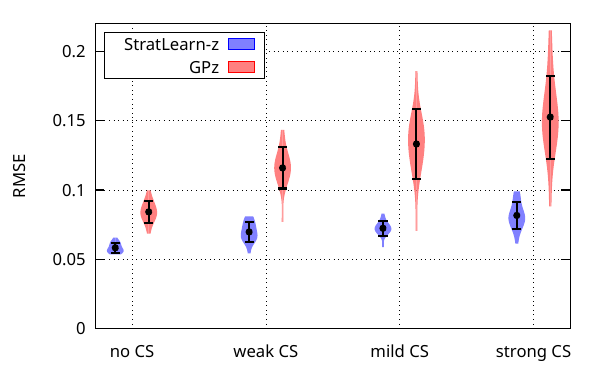}
\includegraphics[width=0.5\textwidth]{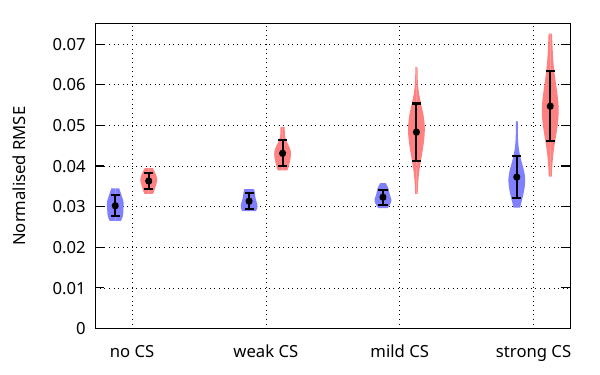}
\includegraphics[width=0.5\textwidth]{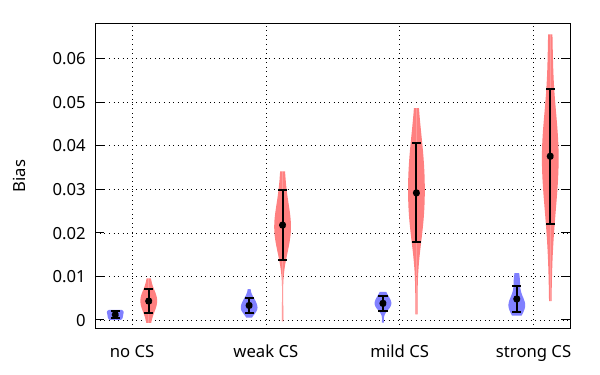}
\includegraphics[width=0.5\textwidth]{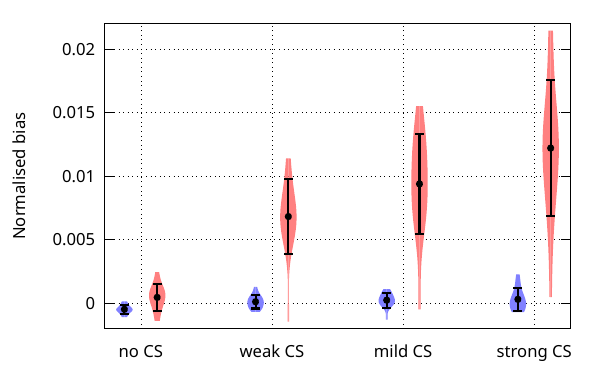}
\includegraphics[width=0.5\textwidth]{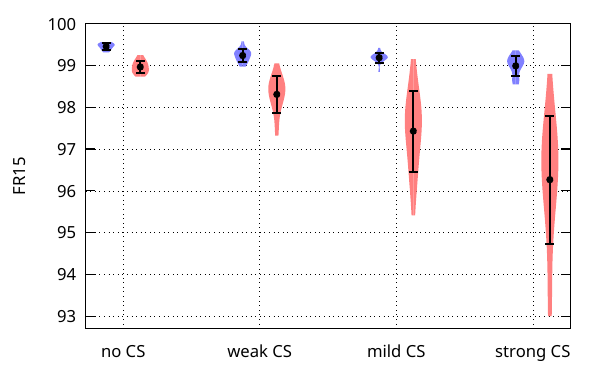}
\includegraphics[width=0.5\textwidth]{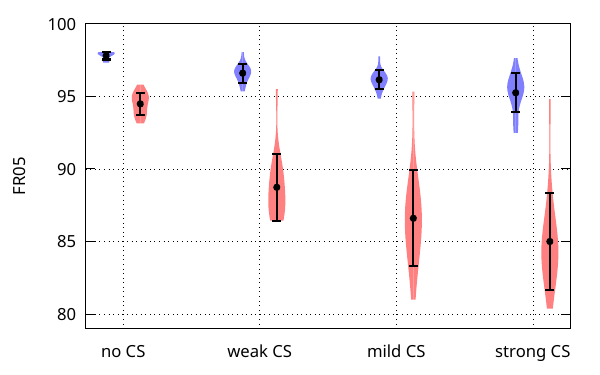}
\caption{Violin plots showing the distribution of the metrics (averaged over 15 data realisations) for the four covariate shift (CS) scenarios. \SLz results are shown in blue, while GPz results are shown in red. Black points with errorbars represent the mean and standard deviation for each case. There is a clear trend in the GPz results, which are strongly degraded by increasing covariate shift. On the other hand, the \SLz results are only slightly impacted.}
\label{fig:metrics-distributions}
\end{figure*}

We perform a more quantitative comparison of the redshift point-estimates by means of the performance metrics described in Sec.~\ref{sec:point-metrics}. The metrics are computed separately for each of the 15 resampled datasets and then averaged. Numerical results for the average performance metrics and standard deviations are collected in Table~\ref{tab:metrics} and illustrated in Fig.~\ref{fig:metrics-distributions}. 

\begin{table*}
  \caption{Average performance metrics and the corresponding standard deviations for \SLz and GPz for the four covariate shift (CS) scenarios considered.}
  \label{tab:metrics}
\centering
\resizebox{0.97\linewidth}{!}{%
\begin{tabular}{lllllllll} 
\hline\hline
                & \multicolumn{2}{c}{no CS} & \multicolumn{2}{c}{weak CS} & \multicolumn{2}{c}{mild CS} & \multicolumn{2}{c}{strong CS}\\ 
\cline{2-9}
                & \SLz & GPz                 & \SLz & GPz                   & \SLz & GPz                   & \SLz & GPz                    \\ 
\hline
RMSE            & $0.0583  \pm 0.0036$ & $0.0842 \pm 0.0079$ & $0.0697 \pm 0.0073$ & $0.1159 \pm 0.0149$ & $0.0724 \pm 0.0054$ & $0.1332 \pm 0.0249$ & $0.0817 \pm 0.0010$ & $0.1526 \pm 0.0299$\\
Norm. RMSE & $0.0302  \pm 0.0025$ & $0.0363 \pm 0.0020$ & $0.0313 \pm 0.0019$ & $0.0431 \pm 0.0032$ & $0.0323 \pm 0.0019$ & $0.0483 \pm 0.0071$ & $0.0373 \pm 0.0052$ & $0.0547 \pm 0.0086$\\
Bias            & $0.0012  \pm 0.0008$ & $0.0043 \pm 0.0027$ & $0.0033 \pm 0.0017$ & $0.0218 \pm 0.0080$ & $0.0038 \pm 0.0017$ & $0.0292 \pm 0.0114$ & $0.0048 \pm 0.0030$ & $0.0376 \pm 0.0155$\\
Norm. bias & $-0.0005 \pm 0.0003$ & $0.0005 \pm 0.0010$ & $0.0001 \pm 0.0005$ & $0.0068 \pm 0.0029$ & $0.0002 \pm 0.0006$ & $0.0094 \pm 0.0039$ & $0.0003 \pm 0.0009$ & $0.0122 \pm 0.0053$\\
FR15            & $99.46   \pm 0.08$   & $98.96  \pm 0.15$   & $99.24  \pm 0.15$   & $98.31  \pm 0.44$   & $99.19  \pm 0.12$   & $97.43  \pm 0.97$   & $98.99  \pm 0.24$   & $96.27  \pm 1.53$  \\
FR05            & $97.81   \pm 0.27$   & $94.48  \pm 0.78$   & $96.59  \pm 0.65$   & $88.73  \pm 2.30$   & $96.14  \pm 0.66$   & $86.59  \pm 3.29$   & $95.25  \pm 1.35$   & $84.99  \pm 3.33$  \\
\hline\hline
\end{tabular}
}
\end{table*}

We note that \SLz outperforms GPz for all metrics and in all cases considered, even in the no covariate shift case, which can likely be ascribed to better performance of the conditional density estimators. However, the difference is more pronounced with more covariate shift.  Both the averaged RMSE and the normalised averaged RMSE, shown in the top panels of Fig.~\ref{fig:metrics-distributions}, show a similar trend: while for GPz they are nearly doubled when going from the case with no covariate shift to the case with strong covariate shift, they are only marginally impacted by the presence of covariate shift in the \SLz results.  Turning to the mean error, or bias (central left panel of Fig.~\ref{fig:metrics-distributions}), the impact of covariate shift on \SLz leads to a mean error that is four times larger in the strong covariate shift scenario with respect to the no covariate shift case, while for GPz the increase is almost one order of magnitude. Moreover, we note that the averaged normalised error for \SLz is always consistent with zero, except for the no covariate shift case where it is slightly lower.
Concerning the FR metric we find again that with \SLz FR is consistent across covariate shift scenarios, with the FR15 (FR05) being only 0.5\% (2.5\%) worse in the strong covariate shift case with respect to the no covariate shift case. For GPz on the other hand FR15 decreases from $\sim99$ in the no covariate shift case to $\sim 96$ in the strong covariate shift case. Similarly, the FR05 ranges from $\sim 98$ to $\sim 95$ for \SLznospace, and from $\sim 95.5$ to $\sim 85$ for GPz.

\begin{figure*}
  \includegraphics[width=\textwidth]{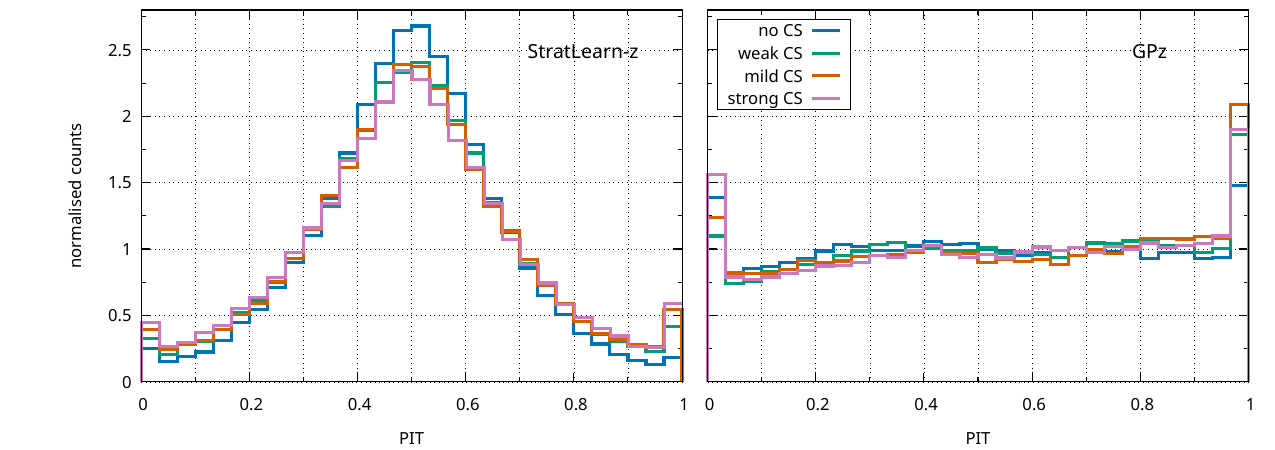}
  \caption{PIT distribution for \SLz (left) and GPz (right) for all cases, as detailed in the legend: in blue the no covariate shift (CS) case, in green the weak covariate shift case, in orange the mild covariate shift case and in green the strong covariate shift case.}
  \label{fig:pit}
\end{figure*}

To evaluate the predicted conditional redshift distributions, Fig.~\ref{fig:pit} plots the histogram of PIT obtained from \SLz and GPz for the four scenarios, focusing on one specific realisation. We checked that the PIT histograms are not significantly different across the 15 realisations. There is an accumulation of mass near 0.5 in the PIT distribution obtained from \SLznospace, indicating an over-estimation of uncertainty in the conditional density estimates. This may prove to be an issue when high-precision estimates are needed, however, we note that the peak is centered at 0.5 and the distributions are symmetric, which suggests that most of the \SLz conditional density estimates are well-centered around the true redshift values. Similar considerations hold for all covariate shift cases, highlighting once again the robustness of \SLz results in presence of unrepresentative source datasets. On the other hand, the PIT distributions obtained from GPz are significantly flatter, but progressively less symmetric the stronger the covariate shift. 

The mode in the PIT distribution of \SLznospace\ indicated that our conditional redshift distribution estimates are too broad, i.e.,  too conservative (see also Appendix \ref{app:pit-toy-model} for a toy model demonstration). 
To better understand this result, we inspected the PIT distribution obtained by replacing our conditional density estimators with GPz, i.e., by running GPz on the stratified data. The resulting PIT distributions are very similar to those obtained from GPz run on the unstratified data (i.e., the distribution shown in the right panel of Fig.~\ref{fig:pit}), suggesting that the mode in the PIT obtained from \SLz can be attributed to the conditional density estimators rather than the stratification. In future work, a possible improvement of the overall \SLz pipeline could be to add GPz to our framework as an additional estimator to potentially flatten the PITs. 
We note that the covariate shirt scenarios considered here are relatively simple and performance as measured by PIT may vary with more  
realistic covariate shift scenarios.

\begin{table}[t]
    \caption{Average of the median CRPS and its standard deviation over 15 realisations for the four covariate shift (CS) scenarios considered, comparing \SLz and GPz (a smaller value indicates better performance).}
    \label{tab:crps}
    \centering
    \begin{tabular}{lcc}
        \hline \hline
         & \SLz & GPz \\ \hline
        no CS          & $0.0079 \pm 0.0001$ & $0.0069 \pm 0.0005$\\ 
        weak CS        & $0.0101 \pm 0.0006$ & $0.0092 \pm 0.0012$\\ 
        mild CS        & $0.0105 \pm 0.0007$ & $0.0093 \pm 0.0013$\\ 
        strong CS      & $0.0112 \pm 0.0010$ & $0.0092 \pm 0.0011$\\ 
        \hline \hline
    \end{tabular}
\end{table}

\begin{figure*}
\includegraphics[width=0.5\textwidth]{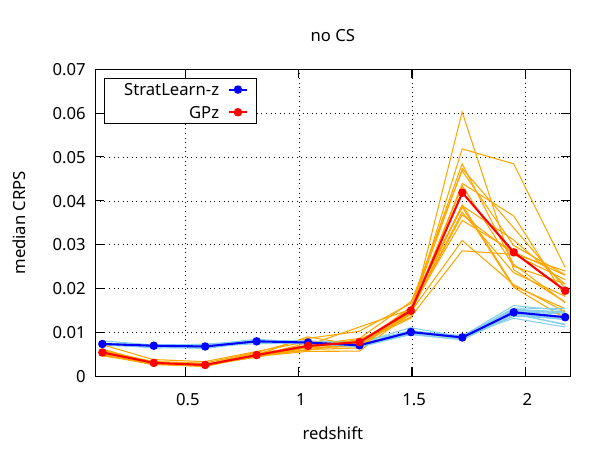}
\includegraphics[width=0.5\textwidth]{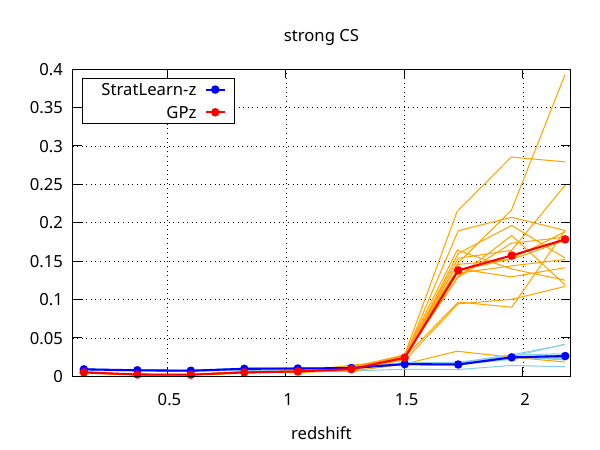}
    \caption{Binned CRPS as a function of redshift for no covariate shift (left) and strong covariate shift (right). Thin orange and light blue lines show the results for each realisation for GPz and \SLz respectively, while thick lines with points represent the average over all realisations.}
\label{fig:binned-crps}
\end{figure*}

Tab.~\ref{tab:crps} reports the median CRPS results for all four scenarios, both for \SLz and GPz, averaged over all 15 realizations.
GPz slightly outperforms \SLz in all cases considered, which is to be expected given that GPz performs better than \SLz as measured by PIT. However, we expect that the ability of \SLz to better predict the mode of conditional redshift distributions (as reflected by its better performance in terms of the point estimate metrics reported in Table \ref{tab:metrics} and Fig.~\ref{fig:metrics-distributions}) should also be reflected in the CRPS, particularly at higher redshifts. To investigate the relative performance of the two methods in different redshift ranges, we computed the CRPS in ten redshift bins by taking the median CRPS in each bin (for each realisation), and then averaging over the realisations in each bin. The results are shown in Fig.~\ref{fig:binned-crps} for the two extreme cases of no covariate shift (left) and strong covariate shift (right). Both cases show similar behavior: in low redshift bins GPz outperforms \SLz due to its tighter variance in the predicted redshift distributions. However, at higher redshifts \SLz consistently achieves a much lower CRPS, thanks to its improved prediction of the mode. The lower overall (i.e., across all redshifts) median CRPS of GPz is thus explained as the result of there being many more lower redshift objects (where GPz outperforms \SLznospace) than higher redshift ones (where \SLz vastly outperforms GPs). We also note that the CRPS obtained from \SLz is roughly consistent across all redshifts, and shows a reduced dependence on the specific realisation than GPz, as demonstrated by \SLz's smaller variance across realizations.
%-------------------------------------------------------------------------------
\section{Conclusions}
\label{sec:conclusion}

We assess the performance of \SLznospace, an implementation of the \SL method proposed in \cite{autenrieth2024a} aimed at photo-$z$ estimation. The method relies on stratification of the source and target datasets based on estimated propensity scores, and is designed to improve the performance of ML algorithms for non-representative training datasets, i.e., in presence of covariate shift. We apply \SLz to a simulated dataset of 100,000 galaxies \citep{derose2019}, where we enforce the presence of covariate shift by performing rejection sampling on the $r$-band magnitudes, following a similar approach as that of \cite{freeman2017}: this prescription is designed to approximately mimic observational bias in spectroscopic (source) datasets, which are biased towards brighter objects with respect to photometric (target) dataset. We vary the parameters involved in the rejection sampling procedure, thus obtaining four different source-target scenarios, with increasing levels of covariate shift. We then estimate propensity scores using logistic regression, and fit two full conditional density estimators to the resulting five strata, combining them via a weighted average. With our trained models we finally obtain predictions for the conditional redshift distributions of objects in the target datasets, which includes brighter and higher redshift objects with respect to the source set.

To evaluate the performance of \SLznospace, we compute a number of metrics on the target predictions, and benchmark against results obtained from the GPz code \citep{almosallam2016a, almosallam2016b}, using the version released with \cite{stylianou2022}.  To obtain more robust estimates for the metrics and investigate their scatter, we randomly permute the full dataset and perform rejection sampling and model training 15 times, thus obtaining 15 sets of photo-$z$ predictions (for each covariate shift scenario) from which we compute averaged metrics. We find \SLz outperforms GPz for all metrics considered and for all degrees of covariate shift.  Additionally, the \SLz results are only marginally impacted by the unrepresentative training data, while GPz results are significantly degraded.

To evaluate the quality of the predicted conditional redshift distribution we plot the histogram of the PIT, for each covariate shift scenario and for both codes. While \SLz overestimates the spread of the conditional redshift distribution, its PIT distribution shows a reduced fraction of catastrophic outliers and symmetry, with a large bump around 0.5, i.e., the \SLz conditional densities appear well-centered around the true redshift values. Despite the overestimation of photo-$z$ pdf spread, we note that the PIT distribution for \SLz is also consistent across covariate shift scenarios, highlighting again the robustness of the method. The GPz histograms are, however, more uniform than the \SLz histograms, which suggests that incorporating GPz or other conditional density estimators within our \SLz framework could lead to improved results.
Additionally, we assessed the conditional redshift distributions via the CRPS metrics, which is informative on both on the variance of the redshift pdf and its mode with respect to the true redshift $\zspec$. We find that, when comparing median CRPS computed over the whole redshift range, GPz slightly outperforms \SLz for all covariate shift scenarios. However, when comparing the CRPS binned as a function of redshift, we showed that \SLz consistently achieves lower values in the high redshift range, indicating once again that an optimal combination of conditional density estimators with our propensity score-based stratification procedure can further improve these results.

The fundamental assumption of this work is that the covariate shift condition holds true, i.e., that there are no unmeasured (confounding) covariates that are predictive for redshift, and also associated to the selection process into source/target set. Covariate shift is often assumed when estimating photo-$z$, e.g, \citep{izbicki2017photo,freeman2017},  but may not be completely accurate for real data. 
This is because the probability of obtaining a spectroscopic redshift may in fact depend on additional factors such as the type of object (e.g., passive galaxies are less likely to deliver a secure redshift compared to emission-line galaxies). Moreover, certain redshift ranges (such as the so-called `redshift desert' $1.4 \leq z \leq 2.5$) consistently lack reference redshifts, which occurs when the [O II] emission line shifts out of the optical range at approximately 0.9 µm, leaving a gap until strong UV lines redshift into the optical domain. 
Furthermore, spectroscopic redshifts measurements can in some cases deviate substantially from the true values, a situation known as `redshift failure'~ \citep{hartley2020}.
To minimize the impact on subsequent analyses of poor quality redshifts, quality flags based on spectral features (e.g., S/N of emission and absorption lines, and the strength of the 4000 Å break) are used to identify and exclude unreliable redshifts \citep{hartley2020}. 
Of course, such spectral features are only observed for spectroscopic confirmed galaxies and usually not available for galaxies in the photometric set, so they cannot be used for stratification. As discussed in \cite{autenrieth2024b}, the covariate shift condition remains valid if the spectroscopic covariates used for quality flagging do not provide additional information about redshift beyond what is already captured by the photometric covariates. 
Since the detailed procedure of obtaining secure redshift, as well as the definition of quality cuts are survey-specific, the definition and application of a general approach to this question is challenging. 
We will investigate potential violations of the covariate shift condition as well as potential correction strategies in a dedicated future study, focusing on topical and realistic survey selection procedures, e.g., as applied for Euclid \citep{laureijs2011, euclidoverview2024}.

In summary, we found the propernsity-score based stratification procedure implemented in \SLz provides stable predictions even in the most extreme scenario considered, in particular for high redshift regions where source/training data are lacking. 
Our analysis shows promising results, specifically concerning applications of the method to current and upcoming galaxy surveys, that will have to face the problem posed by incomplete and/or unrepresentative spectroscopic source sets to maximise the exploitation of photometric data.

%%%%%%%%%%%%%%%%%%%%%%%%%%%%%%%%%%%%%%%%%%%%%%%%%%

\section*{}
{\bf Acknowledgements:} CM's work is supported by the Fondazione ICSC, Spoke 3 Astrophysics and Cosmos Observations, National Recovery and Resilience Plan (Piano Nazionale di Ripresa e Resilienza, PNRR) Project ID CN\_00000013 ``Italian Research Center on High-Performance Computing, Big Data and Quantum Computing'' funded by
MUR Missione 4 Componente 2 Investimento 1.4: Potenziamento strutture di ricerca e creazione di "campioni nazionali di R\&S (M4C2-19 )" - Next Generation EU (NGEU). RT acknowledges co-funding from Next Generation EU, in the context of the National Recovery and Resilience Plan, Investment PE1 – Project FAIR ``Future Artificial Intelligence Research''. This resource was co-financed by the Next Generation EU [DM 1555 del 11.10.22]. RT is partially supported by the Fondazione ICSC, Spoke 3 ``Astrophysics and Cosmos Observations'', Piano Nazionale di Ripresa e Resilienza Project ID CN00000013 ``Italian Research Center on High-Performance Computing, Big Data and Quantum Computing'' funded by MUR Missione 4 Componente 2 Investimento 1.4: Potenziamento strutture di ricerca e creazione di ``campioni nazionali di R\&S (M4C2-19 )'' - Next Generation EU (NGEU). 
DvD and MA acknowledge partial support from the UK Engineering and Physical Sciences Research Council [EP/W015080/1, EP/W522673/1] and  from the Marie-Skodowska-Curie RISE [H2020-MSCA-RISE-2019-873089] Grant provided by the European Commission.
MA also acknowledges partial support from the European Union’s Horizon 2020 research and innovation programme under European Research Council Grant Agreement No 101002652 (PI K. Mandel) and Marie Skłodowska-Curie Grant Agreement No 873089. 
RT and AM have been partially supported by the PRO3 Scuole Programme `DS4ASTRO'. This work used the \href{https://julialang.org/}{Julia} language and in particular the \href{https://dataframes.juliadata.org/stable/}{DataFrames.jl}, \href{https://csv.juliadata.org/stable/}{CSV.jl},  \href{https://docs.julialang.org/en/v1/stdlib/Statistics/}{Statistics.jl}, \href{https://github.com/JuliaStats/Distances.jl}{Distances}, \href{https://juliastats.org/GLM.jl/stable/}{GLM.jl} packages, and the \href{https://gcalderone.github.io/Gnuplot.jl/stable/index.html}{Gnuplot.jl} package for Julia to generate the plots.

%%%%%%%%%%%%%%%%%%%%%%%%%%%%%%%%%%%%%%%%%%%%%%%%%%
\section*{Data Availability}

The original dataset used in this paper can be found at \url{https://github.com/jfcrenshaw/pzflow/blob/main/pzflow/example_files/galaxy-data.pkl}.
We publicly release the version of \SL used for this work with example datasets -- one for each covariate shift case considered -- at \href{https://github.com/chiaramoretti/StratLearn-z}{github.com/chiaramoretti/StratLearn-z}



\bibliographystyle{mnras}
\bibliography{main} % if your bibtex file is called example.bib

\begin{thebibliography}{}
\makeatletter
\relax
\def\mn@urlcharsother{\let\do\@makeother \do\$\do\&\do\#\do\^\do\_\do\%\do\~}
\def\mn@doi{\begingroup\mn@urlcharsother \@ifnextchar [ {\mn@doi@}
  {\mn@doi@[]}}
\def\mn@doi@[#1]#2{\def\@tempa{#1}\ifx\@tempa\@empty \href
  {http://dx.doi.org/#2} {doi:#2}\else \href {http://dx.doi.org/#2} {#1}\fi
  \endgroup}
\def\mn@eprint#1#2{\mn@eprint@#1:#2::\@nil}
\def\mn@eprint@arXiv#1{\href {http://arxiv.org/abs/#1} {{\tt arXiv:#1}}}
\def\mn@eprint@dblp#1{\href {http://dblp.uni-trier.de/rec/bibtex/#1.xml}
  {dblp:#1}}
\def\mn@eprint@#1:#2:#3:#4\@nil{\def\@tempa {#1}\def\@tempb {#2}\def\@tempc
  {#3}\ifx \@tempc \@empty \let \@tempc \@tempb \let \@tempb \@tempa \fi \ifx
  \@tempb \@empty \def\@tempb {arXiv}\fi \@ifundefined
  {mn@eprint@\@tempb}{\@tempb:\@tempc}{\expandafter \expandafter \csname
  mn@eprint@\@tempb\endcsname \expandafter{\@tempc}}}

\bibitem[\protect\citeauthoryear{{Abbott} et~al.,}{{Abbott}
  et~al.}{2018}]{abbott2018}
{Abbott} T.~M.~C.,  et~al., 2018, \mn@doi [\apjs] {10.3847/1538-4365/aae9f0},
  \href {https://ui.adsabs.harvard.edu/abs/2018ApJS..239...18A} {239, 18}

\bibitem[\protect\citeauthoryear{{Aihara} et~al.,}{{Aihara}
  et~al.}{2018}]{aihara2018}
{Aihara} H.,  et~al., 2018, \mn@doi [\pasj] {10.1093/pasj/psx081}, \href
  {https://ui.adsabs.harvard.edu/abs/2018PASJ...70S...8A} {70, S8}

\bibitem[\protect\citeauthoryear{{Akeson} et~al.,}{{Akeson}
  et~al.}{2019}]{akeson2019}
{Akeson} R.,  et~al., 2019, \mn@doi [arXiv e-prints]
  {10.48550/arXiv.1902.05569}, \href
  {https://ui.adsabs.harvard.edu/abs/2019arXiv190205569A} {p. arXiv:1902.05569}

\bibitem[\protect\citeauthoryear{{Almosallam}, {Lindsay}, {Jarvis}  \&
  {Roberts}}{{Almosallam} et~al.}{2016a}]{almosallam2016a}
{Almosallam} I.~A.,  {Lindsay} S.~N.,  {Jarvis} M.~J.,   {Roberts} S.~J.,
  2016a, \mn@doi [\mnras] {10.1093/mnras/stv2425}, \href
  {https://ui.adsabs.harvard.edu/abs/2016MNRAS.455.2387A} {455, 2387}

\bibitem[\protect\citeauthoryear{{Almosallam}, {Jarvis}  \&
  {Roberts}}{{Almosallam} et~al.}{2016b}]{almosallam2016b}
{Almosallam} I.~A.,  {Jarvis} M.~J.,   {Roberts} S.~J.,  2016b, \mn@doi
  [\mnras] {10.1093/mnras/stw1618}, \href
  {https://ui.adsabs.harvard.edu/abs/2016MNRAS.462..726A} {462, 726}

\bibitem[\protect\citeauthoryear{{Amara} \& {R{\'e}fr{\'e}gier}}{{Amara} \&
  {R{\'e}fr{\'e}gier}}{2007}]{amara2007}
{Amara} A.,  {R{\'e}fr{\'e}gier} A.,  2007, \mn@doi [\mnras]
  {10.1111/j.1365-2966.2007.12271.x}, \href
  {https://ui.adsabs.harvard.edu/abs/2007MNRAS.381.1018A} {381, 1018}

\bibitem[\protect\citeauthoryear{{Arnouts}, {Cristiani}, {Moscardini},
  {Matarrese}, {Lucchin}, {Fontana}  \& {Giallongo}}{{Arnouts}
  et~al.}{1999}]{arnouts1999}
{Arnouts} S.,  {Cristiani} S.,  {Moscardini} L.,  {Matarrese} S.,  {Lucchin}
  F.,  {Fontana} A.,   {Giallongo} E.,  1999, \mn@doi [\mnras]
  {10.1046/j.1365-8711.1999.02978.x}, \href
  {https://ui.adsabs.harvard.edu/abs/1999MNRAS.310..540A} {310, 540}

\bibitem[\protect\citeauthoryear{{Autenrieth}, {Wright}, {Trotta}, {van Dyk},
  {Stenning}  \& {Joachimi}}{{Autenrieth} et~al.}{2024a}]{autenrieth2024b}
{Autenrieth} M.,  {Wright} A.~H.,  {Trotta} R.,  {van Dyk} D.~A.,  {Stenning}
  D.~C.,   {Joachimi} B.,  2024a, \mn@doi [arXiv e-prints]
  {10.48550/arXiv.2401.04687}, \href
  {https://ui.adsabs.harvard.edu/abs/2024arXiv240104687A} {p. arXiv:2401.04687}

\bibitem[\protect\citeauthoryear{Autenrieth, van Dyk, Trotta  \&
  Stenning}{Autenrieth et~al.}{2024b}]{autenrieth2024a}
Autenrieth M.,  van Dyk D.~A.,  Trotta R.,   Stenning D.~C.,  2024b, \mn@doi
  [Statistical Analysis and Data Mining: The ASA Data Science Journal]
  {https://doi.org/10.1002/sam.11643}, 17, e11643

\bibitem[\protect\citeauthoryear{{Ben{\'\i}tez}}{{Ben{\'\i}tez}}{2000}]{benitez2000}
{Ben{\'\i}tez} N.,  2000, \mn@doi [\apj] {10.1086/308947}, \href
  {https://ui.adsabs.harvard.edu/abs/2000ApJ...536..571B} {536, 571}

\bibitem[\protect\citeauthoryear{{Bolzonella}, {Miralles}  \&
  {Pell{\'o}}}{{Bolzonella} et~al.}{2000}]{bolzonella2000}
{Bolzonella} M.,  {Miralles} J.~M.,   {Pell{\'o}} R.,  2000, \mn@doi [\aap]
  {10.48550/arXiv.astro-ph/0003380}, \href
  {https://ui.adsabs.harvard.edu/abs/2000A&A...363..476B} {363, 476}

\bibitem[\protect\citeauthoryear{{Bordoloi}, {Lilly}  \& {Amara}}{{Bordoloi}
  et~al.}{2010}]{bordoloi2010}
{Bordoloi} R.,  {Lilly} S.~J.,   {Amara} A.,  2010, \mn@doi [\mnras]
  {10.1111/j.1365-2966.2010.16765.x}, \href
  {https://ui.adsabs.harvard.edu/abs/2010MNRAS.406..881B} {406, 881}

\bibitem[\protect\citeauthoryear{{Brammer}, {van Dokkum}  \& {Coppi}}{{Brammer}
  et~al.}{2008}]{brammer2008}
{Brammer} G.~B.,  {van Dokkum} P.~G.,   {Coppi} P.,  2008, \mn@doi [\apj]
  {10.1086/591786}, \href
  {https://ui.adsabs.harvard.edu/abs/2008ApJ...686.1503B} {686, 1503}

\bibitem[\protect\citeauthoryear{{Brescia}, {Cavuoti}, {Razim}, {Amaro},
  {Riccio}  \& {Longo}}{{Brescia} et~al.}{2021}]{brescia2021}
{Brescia} M.,  {Cavuoti} S.,  {Razim} O.,  {Amaro} V.,  {Riccio} G.,   {Longo}
  G.,  2021, \mn@doi [Frontiers in Astronomy and Space Sciences]
  {10.3389/fspas.2021.658229}, \href
  {https://ui.adsabs.harvard.edu/abs/2021FrASS...8...70B} {8, 70}

\bibitem[\protect\citeauthoryear{{Carliles}, {Budav{\'a}ri}, {Heinis}, {Priebe}
   \& {Szalay}}{{Carliles} et~al.}{2010}]{carliles2010}
{Carliles} S.,  {Budav{\'a}ri} T.,  {Heinis} S.,  {Priebe} C.,   {Szalay}
  A.~S.,  2010, \mn@doi [\apj] {10.1088/0004-637X/712/1/511}, \href
  {https://ui.adsabs.harvard.edu/abs/2010ApJ...712..511C} {712, 511}

\bibitem[\protect\citeauthoryear{{Carrasco Kind} \& {Brunner}}{{Carrasco Kind}
  \& {Brunner}}{2013}]{carrascokind2013}
{Carrasco Kind} M.,  {Brunner} R.~J.,  2013, \mn@doi [\mnras]
  {10.1093/mnras/stt574}, \href
  {https://ui.adsabs.harvard.edu/abs/2013MNRAS.432.1483C} {432, 1483}

\bibitem[\protect\citeauthoryear{{Carrasco Kind} \& {Brunner}}{{Carrasco Kind}
  \& {Brunner}}{2014}]{carrascokind2014}
{Carrasco Kind} M.,  {Brunner} R.~J.,  2014, \mn@doi [\mnras]
  {10.1093/mnras/stt2456}, \href
  {https://ui.adsabs.harvard.edu/abs/2014MNRAS.438.3409C} {438, 3409}

\bibitem[\protect\citeauthoryear{Cochran}{Cochran}{1968}]{cochran1968}
Cochran W.~G.,  1968, Biometrics, pp 295--313

\bibitem[\protect\citeauthoryear{{Collister} \& {Lahav}}{{Collister} \&
  {Lahav}}{2004}]{collister2004}
{Collister} A.~A.,  {Lahav} O.,  2004, \mn@doi [\pasp] {10.1086/383254}, \href
  {https://ui.adsabs.harvard.edu/abs/2004PASP..116..345C} {116, 345}

\bibitem[\protect\citeauthoryear{{Crenshaw}, {Kalmbach}, {Gagliano}, {Yan},
  {Connolly}, {Malz}, {Schmidt}  \& {The LSST Dark Energy Science
  Collaboration}}{{Crenshaw} et~al.}{2024}]{crenshaw2024}
{Crenshaw} J.~F.,  {Kalmbach} J.~B.,  {Gagliano} A.,  {Yan} Z.,  {Connolly}
  A.~J.,  {Malz} A.~I.,  {Schmidt} S.~J.,   {The LSST Dark Energy Science
  Collaboration} 2024, \mn@doi [\aj] {10.3847/1538-3881/ad54bf}, \href
  {https://ui.adsabs.harvard.edu/abs/2024AJ....168...80C} {168, 80}

\bibitem[\protect\citeauthoryear{{D'Isanto} \& {Polsterer}}{{D'Isanto} \&
  {Polsterer}}{2018}]{disanto2018}
{D'Isanto} A.,  {Polsterer} K.~L.,  2018, \mn@doi [\aap]
  {10.1051/0004-6361/201731326}, \href
  {https://ui.adsabs.harvard.edu/abs/2018A&A...609A.111D} {609, A111}

\bibitem[\protect\citeauthoryear{Dawid}{Dawid}{1984}]{dawid1984}
Dawid A.~P.,  1984, \mn@doi [Journal of the Royal Statistical Society: Series A
  (General)] {https://doi.org/10.2307/2981683}, 147, 278

\bibitem[\protect\citeauthoryear{{DeRose} et~al.,}{{DeRose}
  et~al.}{2019}]{derose2019}
{DeRose} J.,  et~al., 2019, \mn@doi [arXiv e-prints]
  {10.48550/arXiv.1901.02401}, \href
  {https://ui.adsabs.harvard.edu/abs/2019arXiv190102401D} {p. arXiv:1901.02401}

\bibitem[\protect\citeauthoryear{{Euclid Collaboration} et~al.,}{{Euclid
  Collaboration} et~al.}{2020}]{desprez2020}
{Euclid Collaboration} et~al., 2020, \mn@doi [\aap]
  {10.1051/0004-6361/202039403}, \href
  {https://ui.adsabs.harvard.edu/abs/2020A&A...644A..31E} {644, A31}

\bibitem[\protect\citeauthoryear{{Euclid Collaboration} et~al.,}{{Euclid
  Collaboration} et~al.}{2024}]{euclidoverview2024}
{Euclid Collaboration} et~al., 2024, \mn@doi [arXiv e-prints]
  {10.48550/arXiv.2405.13491}, \href
  {https://ui.adsabs.harvard.edu/abs/2024arXiv240513491E} {p. arXiv:2405.13491}

\bibitem[\protect\citeauthoryear{{Firth}, {Lahav}  \& {Somerville}}{{Firth}
  et~al.}{2003}]{firth2003}
{Firth} A.~E.,  {Lahav} O.,   {Somerville} R.~S.,  2003, \mn@doi [\mnras]
  {10.1046/j.1365-8711.2003.06271.x}, \href
  {https://ui.adsabs.harvard.edu/abs/2003MNRAS.339.1195F} {339, 1195}

\bibitem[\protect\citeauthoryear{{Freeman}, {Izbicki}  \& {Lee}}{{Freeman}
  et~al.}{2017}]{freeman2017}
{Freeman} P.~E.,  {Izbicki} R.,   {Lee} A.~B.,  2017, \mn@doi [\mnras]
  {10.1093/mnras/stx764}, \href
  {https://ui.adsabs.harvard.edu/abs/2017MNRAS.468.4556F} {468, 4556}

\bibitem[\protect\citeauthoryear{{Graff}, {Feroz}, {Hobson}  \&
  {Lasenby}}{{Graff} et~al.}{2014}]{graff2014}
{Graff} P.,  {Feroz} F.,  {Hobson} M.~P.,   {Lasenby} A.,  2014, \mn@doi
  [\mnras] {10.1093/mnras/stu642}, \href
  {https://ui.adsabs.harvard.edu/abs/2014MNRAS.441.1741G} {441, 1741}

\bibitem[\protect\citeauthoryear{{Graham}, {Connolly}, {Ivezi{\'c}}, {Schmidt},
  {Jones}, {Juri{\'c}}, {Daniel}  \& {Yoachim}}{{Graham}
  et~al.}{2018}]{graham2018}
{Graham} M.~L.,  {Connolly} A.~J.,  {Ivezi{\'c}} {\v{Z}}.,  {Schmidt} S.~J.,
  {Jones} R.~L.,  {Juri{\'c}} M.,  {Daniel} S.~F.,   {Yoachim} P.,  2018,
  \mn@doi [\aj] {10.3847/1538-3881/aa99d4}, \href
  {https://ui.adsabs.harvard.edu/abs/2018AJ....155....1G} {155, 1}

\bibitem[\protect\citeauthoryear{{Hartley} et~al.,}{{Hartley}
  et~al.}{2020}]{hartley2020}
{Hartley} W.~G.,  et~al., 2020, \mn@doi [\mnras] {10.1093/mnras/staa1812},
  \href {https://ui.adsabs.harvard.edu/abs/2020MNRAS.496.4769H} {496, 4769}

\bibitem[\protect\citeauthoryear{{Henghes}, {Thiyagalingam}, {Pettitt}, {Hey}
  \& {Lahav}}{{Henghes} et~al.}{2022}]{henghes2022}
{Henghes} B.,  {Thiyagalingam} J.,  {Pettitt} C.,  {Hey} T.,   {Lahav} O.,
  2022, \mn@doi [\mnras] {10.1093/mnras/stac480}, \href
  {https://ui.adsabs.harvard.edu/abs/2022MNRAS.512.1696H} {512, 1696}

\bibitem[\protect\citeauthoryear{{Hersbach}}{{Hersbach}}{2000}]{hersbach2000}
{Hersbach} H.,  2000, \mn@doi [Weather and Forecasting]
  {10.1175/1520-0434(2000)015<0559:DOTCRP>2.0.CO;2}, \href
  {https://ui.adsabs.harvard.edu/abs/2000WtFor..15..559H} {15, 559}

\bibitem[\protect\citeauthoryear{{Heymans} et~al.,}{{Heymans}
  et~al.}{2021}]{heymans2021}
{Heymans} C.,  et~al., 2021, \mn@doi [\aap] {10.1051/0004-6361/202039063},
  \href {https://ui.adsabs.harvard.edu/abs/2021A&A...646A.140H} {646, A140}

\bibitem[\protect\citeauthoryear{{Hildebrandt} et~al.,}{{Hildebrandt}
  et~al.}{2010}]{hildebrandt2010}
{Hildebrandt} H.,  et~al., 2010, \mn@doi [\aap] {10.1051/0004-6361/201014885},
  \href {https://ui.adsabs.harvard.edu/abs/2010A&A...523A..31H} {523, A31}

\bibitem[\protect\citeauthoryear{{Hildebrandt} et~al.,}{{Hildebrandt}
  et~al.}{2021}]{hildebrandt2021}
{Hildebrandt} H.,  et~al., 2021, \mn@doi [\aap] {10.1051/0004-6361/202039018},
  \href {https://ui.adsabs.harvard.edu/abs/2021A&A...647A.124H} {647, A124}

\bibitem[\protect\citeauthoryear{{Ilbert} et~al.,}{{Ilbert}
  et~al.}{2006}]{ilbert2006}
{Ilbert} O.,  et~al., 2006, \mn@doi [\aap] {10.1051/0004-6361:20065138}, \href
  {https://ui.adsabs.harvard.edu/abs/2006A&A...457..841I} {457, 841}

\bibitem[\protect\citeauthoryear{{Ivezi{\'c}} et~al.,}{{Ivezi{\'c}}
  et~al.}{2019}]{ivezic2019}
{Ivezi{\'c}} {\v{Z}}.,  et~al., 2019, \mn@doi [\apj]
  {10.3847/1538-4357/ab042c}, \href
  {https://ui.adsabs.harvard.edu/abs/2019ApJ...873..111I} {873, 111}

\bibitem[\protect\citeauthoryear{{Izbicki}, {Lee}  \& {Freeman}}{{Izbicki}
  et~al.}{2016}]{izbicki2016}
{Izbicki} R.,  {Lee} A.~B.,   {Freeman} P.~E.,  2016, \mn@doi [arXiv e-prints]
  {10.48550/arXiv.1604.01339}, \href
  {https://ui.adsabs.harvard.edu/abs/2016arXiv160401339I} {p. arXiv:1604.01339}

\bibitem[\protect\citeauthoryear{Izbicki, Lee, Freeman  et~al.}{Izbicki
  et~al.}{2017}]{izbicki2017photo}
Izbicki R.,  Lee A.~B.,  Freeman P.~E.,   et~al., 2017, The Annals of Applied
  Statistics, 11, 698

\bibitem[\protect\citeauthoryear{{Laureijs} et~al.,}{{Laureijs}
  et~al.}{2011}]{laureijs2011}
{Laureijs} R.,  et~al., 2011, \mn@doi [arXiv e-prints]
  {10.48550/arXiv.1110.3193}, \href
  {https://ui.adsabs.harvard.edu/abs/2011arXiv1110.3193L} {p. arXiv:1110.3193}

\bibitem[\protect\citeauthoryear{{Ma}, {Hu}  \& {Huterer}}{{Ma}
  et~al.}{2006}]{ma2006}
{Ma} Z.,  {Hu} W.,   {Huterer} D.,  2006, \mn@doi [\apj] {10.1086/497068},
  \href {https://ui.adsabs.harvard.edu/abs/2006ApJ...636...21M} {636, 21}

\bibitem[\protect\citeauthoryear{{Malmquist}}{{Malmquist}}{1922}]{malmquist1922}
{Malmquist} K.~G.,  1922, Meddelanden fran Lunds Astronomiska Observatorium
  Serie I, \href {https://ui.adsabs.harvard.edu/abs/1922MeLuF.100....1M} {100,
  1}

\bibitem[\protect\citeauthoryear{{Malmquist}}{{Malmquist}}{1925}]{malmquist1925}
{Malmquist} K.~G.,  1925, Meddelanden fran Lunds Astronomiska Observatorium
  Serie I, \href {https://ui.adsabs.harvard.edu/abs/1925MeLuF.106....1M} {106,
  1}

\bibitem[\protect\citeauthoryear{{Masters} et~al.,}{{Masters}
  et~al.}{2015}]{masters2015}
{Masters} D.,  et~al., 2015, \mn@doi [\apj] {10.1088/0004-637X/813/1/53}, \href
  {https://ui.adsabs.harvard.edu/abs/2015ApJ...813...53M} {813, 53}

\bibitem[\protect\citeauthoryear{{Moreno-Torres}, {Raeder},
  {Alaiz-Rodr{\'\i}guez}, {Chawla}  \& {Herrera}}{{Moreno-Torres}
  et~al.}{2012}]{moreno-torres2012}
{Moreno-Torres} J.~G.,  {Raeder} T.,  {Alaiz-Rodr{\'\i}guez} R.,  {Chawla}
  N.~V.,   {Herrera} F.,  2012, \mn@doi [Pattern Recognition]
  {10.1016/j.patcog.2011.06.019}, \href
  {https://ui.adsabs.harvard.edu/abs/2012PatRe..45..521M} {45, 521}

\bibitem[\protect\citeauthoryear{{Moskowitz}, {Gawiser}, {Crenshaw}, {Andrews},
  {Schmidt}  \& {The LSST Dark Energy Science Collaboration}}{{Moskowitz}
  et~al.}{2024}]{moskowitz2024}
{Moskowitz} I.,  {Gawiser} E.,  {Crenshaw} J.~F.,  {Andrews} B.~H.,  {Schmidt}
  S.,   {The LSST Dark Energy Science Collaboration} 2024, \mn@doi [arXiv
  e-prints] {10.48550/arXiv.2402.15551}, \href
  {https://ui.adsabs.harvard.edu/abs/2024arXiv240215551M} {p. arXiv:2402.15551}

\bibitem[\protect\citeauthoryear{{Myles} et~al.,}{{Myles}
  et~al.}{2021}]{myles2021}
{Myles} J.,  et~al., 2021, \mn@doi [\mnras] {10.1093/mnras/stab1515}, \href
  {https://ui.adsabs.harvard.edu/abs/2021MNRAS.505.4249M} {505, 4249}

\bibitem[\protect\citeauthoryear{{Newman} \& {Gruen}}{{Newman} \&
  {Gruen}}{2022}]{newman2022}
{Newman} J.~A.,  {Gruen} D.,  2022, \mn@doi [\araa]
  {10.1146/annurev-astro-032122-014611}, \href
  {https://ui.adsabs.harvard.edu/abs/2022ARA&A..60..363N} {60, 363}

\bibitem[\protect\citeauthoryear{{Pasquet}, {Bertin}, {Treyer}, {Arnouts}  \&
  {Fouchez}}{{Pasquet} et~al.}{2019}]{pasquet2019}
{Pasquet} J.,  {Bertin} E.,  {Treyer} M.,  {Arnouts} S.,   {Fouchez} D.,  2019,
  \mn@doi [\aap] {10.1051/0004-6361/201833617}, \href
  {https://ui.adsabs.harvard.edu/abs/2019A&A...621A..26P} {621, A26}

\bibitem[\protect\citeauthoryear{{Polsterer}, {D'Isanto}  \&
  {Gieseke}}{{Polsterer} et~al.}{2016}]{polsterer2016}
{Polsterer} K.~L.,  {D'Isanto} A.,   {Gieseke} F.,  2016, \mn@doi [arXiv
  e-prints] {10.48550/arXiv.1608.08016}, \href
  {https://ui.adsabs.harvard.edu/abs/2016arXiv160808016P} {p. arXiv:1608.08016}

\bibitem[\protect\citeauthoryear{{Porredon} et~al.,}{{Porredon}
  et~al.}{2022}]{porredon2022}
{Porredon} A.,  et~al., 2022, \mn@doi [\prd] {10.1103/PhysRevD.106.103530},
  \href {https://ui.adsabs.harvard.edu/abs/2022PhRvD.106j3530P} {106, 103530}

\bibitem[\protect\citeauthoryear{{Rau} et~al.,}{{Rau} et~al.}{2023}]{rau2023}
{Rau} M.~M.,  et~al., 2023, \mn@doi [\mnras] {10.1093/mnras/stad1962}, \href
  {https://ui.adsabs.harvard.edu/abs/2023MNRAS.524.5109R} {524, 5109}

\bibitem[\protect\citeauthoryear{Rosenbaum \& Rubin}{Rosenbaum \&
  Rubin}{1983}]{rosenbaum1983}
Rosenbaum P.~R.,  Rubin D.~B.,  1983, Biometrika, 70, 41

\bibitem[\protect\citeauthoryear{{Sadeh}, {Abdalla}  \& {Lahav}}{{Sadeh}
  et~al.}{2016}]{sadeh2016}
{Sadeh} I.,  {Abdalla} F.~B.,   {Lahav} O.,  2016, \mn@doi [\pasp]
  {10.1088/1538-3873/128/968/104502}, \href
  {https://ui.adsabs.harvard.edu/abs/2016PASP..128j4502S} {128, 104502}

\bibitem[\protect\citeauthoryear{{Salvato}, {Ilbert}  \& {Hoyle}}{{Salvato}
  et~al.}{2019}]{salvato2019}
{Salvato} M.,  {Ilbert} O.,   {Hoyle} B.,  2019, \mn@doi [Nature Astronomy]
  {10.1038/s41550-018-0478-0}, \href
  {https://ui.adsabs.harvard.edu/abs/2019NatAs...3..212S} {3, 212}

\bibitem[\protect\citeauthoryear{{S{\'a}nchez} et~al.,}{{S{\'a}nchez}
  et~al.}{2014}]{sanchez2014}
{S{\'a}nchez} C.,  et~al., 2014, \mn@doi [\mnras] {10.1093/mnras/stu1836},
  \href {https://ui.adsabs.harvard.edu/abs/2014MNRAS.445.1482S} {445, 1482}

\bibitem[\protect\citeauthoryear{{Schmidt} et~al.,}{{Schmidt}
  et~al.}{2020}]{schmidt2020}
{Schmidt} S.~J.,  et~al., 2020, \mn@doi [\mnras] {10.1093/mnras/staa2799},
  \href {https://ui.adsabs.harvard.edu/abs/2020MNRAS.499.1587S} {499, 1587}

\bibitem[\protect\citeauthoryear{{Stylianou}, {Malz}, {Hatfield}, {Crenshaw}
  \& {Gschwend}}{{Stylianou} et~al.}{2022}]{stylianou2022}
{Stylianou} N.,  {Malz} A.~I.,  {Hatfield} P.,  {Crenshaw} J.~F.,   {Gschwend}
  J.,  2022, \mn@doi [\pasp] {10.1088/1538-3873/ac59bf}, \href
  {https://ui.adsabs.harvard.edu/abs/2022PASP..134d4501S} {134, 044501}

\bibitem[\protect\citeauthoryear{{Sugiyama} et~al.,}{{Sugiyama}
  et~al.}{2023}]{sugiyama2023}
{Sugiyama} S.,  et~al., 2023, \mn@doi [\prd] {10.1103/PhysRevD.108.123521},
  \href {https://ui.adsabs.harvard.edu/abs/2023PhRvD.108l3521S} {108, 123521}

\bibitem[\protect\citeauthoryear{{Tanaka} et~al.,}{{Tanaka}
  et~al.}{2018}]{tanaka2018}
{Tanaka} M.,  et~al., 2018, \mn@doi [\pasj] {10.1093/pasj/psx077}, \href
  {https://ui.adsabs.harvard.edu/abs/2018PASJ...70S...9T} {70, S9}

\bibitem[\protect\citeauthoryear{{The LSST Dark Energy Science Collaboration}
  et~al.,}{{The LSST Dark Energy Science Collaboration}
  et~al.}{2018}]{lsst2018}
{The LSST Dark Energy Science Collaboration} et~al., 2018, \mn@doi [arXiv
  e-prints] {10.48550/arXiv.1809.01669}, \href
  {https://ui.adsabs.harvard.edu/abs/2018arXiv180901669T} {p. arXiv:1809.01669}

\bibitem[\protect\citeauthoryear{{Toribio San Cipriano} et~al.,}{{Toribio San
  Cipriano} et~al.}{2023}]{sancipriano2023}
{Toribio San Cipriano} L.,  et~al., 2023, \mn@doi [arXiv e-prints]
  {10.48550/arXiv.2312.09721}, \href
  {https://ui.adsabs.harvard.edu/abs/2023arXiv231209721T} {p. arXiv:2312.09721}

\bibitem[\protect\citeauthoryear{{Tosone}, {Cagliari}, {Guzzo}, {Granett}  \&
  {Crespi}}{{Tosone} et~al.}{2023}]{tosone2023}
{Tosone} F.,  {Cagliari} M.~S.,  {Guzzo} L.,  {Granett} B.~R.,   {Crespi} A.,
  2023, \mn@doi [\aap] {10.1051/0004-6361/202245369}, \href
  {https://ui.adsabs.harvard.edu/abs/2023A&A...672A.150T} {672, A150}

\bibitem[\protect\citeauthoryear{{Tutusaus} et~al.,}{{Tutusaus}
  et~al.}{2020}]{tutusaus2020}
{Tutusaus} I.,  et~al., 2020, \mn@doi [\aap] {10.1051/0004-6361/202038313},
  \href {https://ui.adsabs.harvard.edu/abs/2020A&A...643A..70T} {643, A70}

\makeatother
\end{thebibliography}

%%%%%%%%%%%%%%%%% APPENDICES %%%%%%%%%%%%%%%%%%%%%

\appendix

\section{Distributions for all photometric bands}
\label{app:band-distrib}

\begin{figure}[t!]
\centering
\includegraphics[width=0.95\columnwidth]{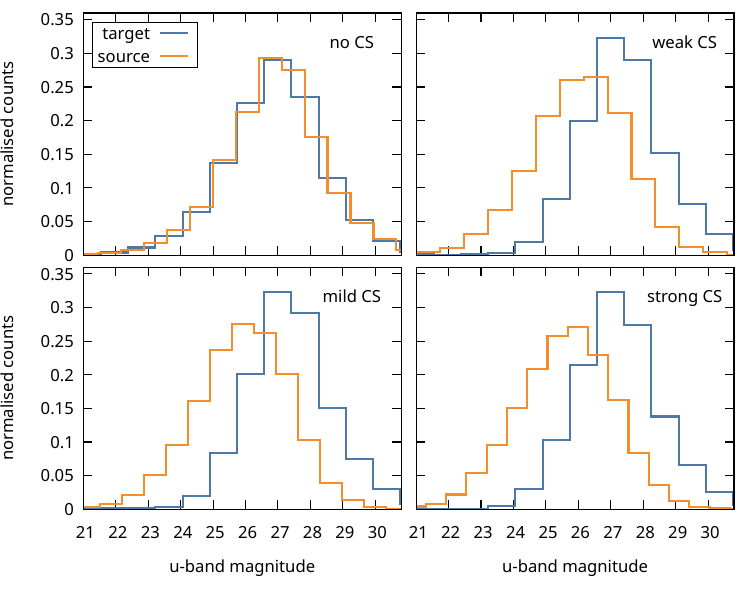}
\caption{Source and target normalised distributions of $u$-band magnitudes for the four covariate shift (CS) cases, as stated in each panel. Orange histograms represent the source dataset, while in blue we plot the target datasets.}
\label{fig:u-band}
\end{figure}

\begin{figure}[t!]
\centering
\includegraphics[width=0.95\columnwidth]{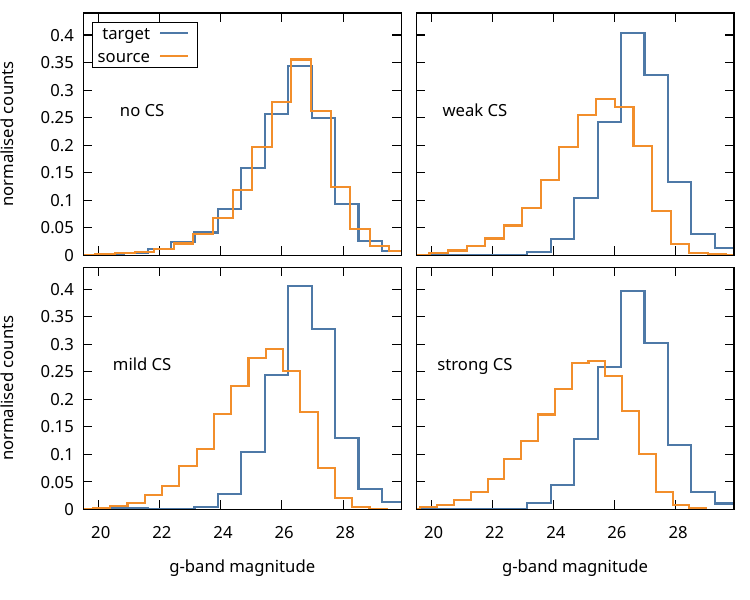}
\caption{Same as Fig.~\ref{fig:u-band} but for the $g$-band magnitudes.}
\label{fig:g-band}
\end{figure}

\begin{figure}[t!]
\centering
\includegraphics[width=0.95\columnwidth]{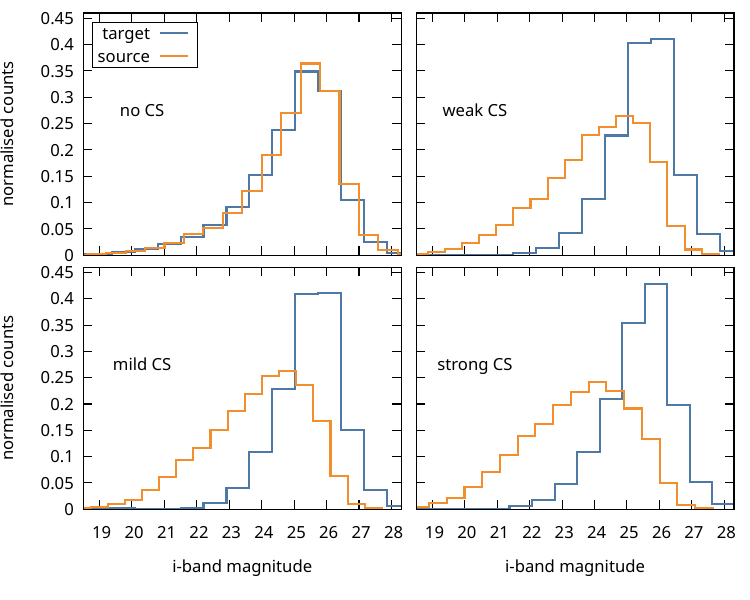}
\caption{Same as Fig.~\ref{fig:u-band} but for the $i$-band magnitudes.}
\label{fig:i-band}
\end{figure}

\begin{figure}[t!]
\centering
\includegraphics[width=0.95\columnwidth]{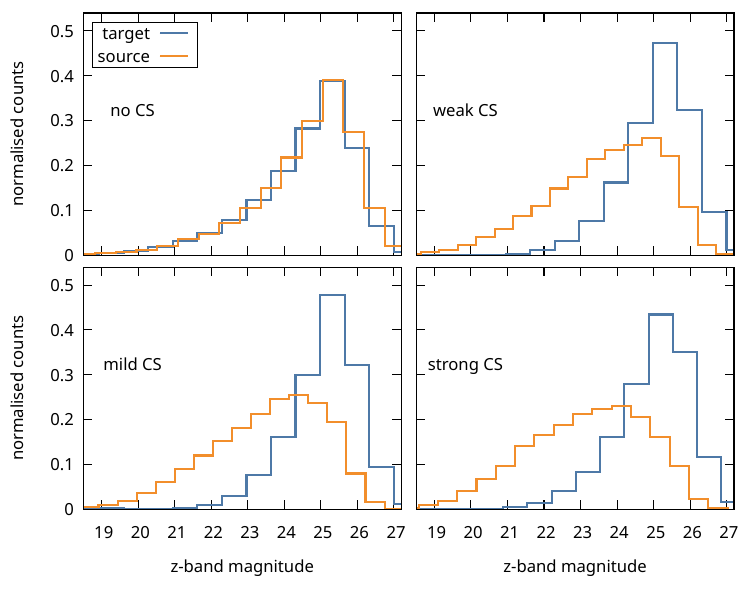}
\caption{Same as Fig.~\ref{fig:u-band} but for the $z$-band magnitudes.}
\label{fig:z-band}
\end{figure}

\begin{figure}[t!]
\centering
\includegraphics[width=0.95\columnwidth]{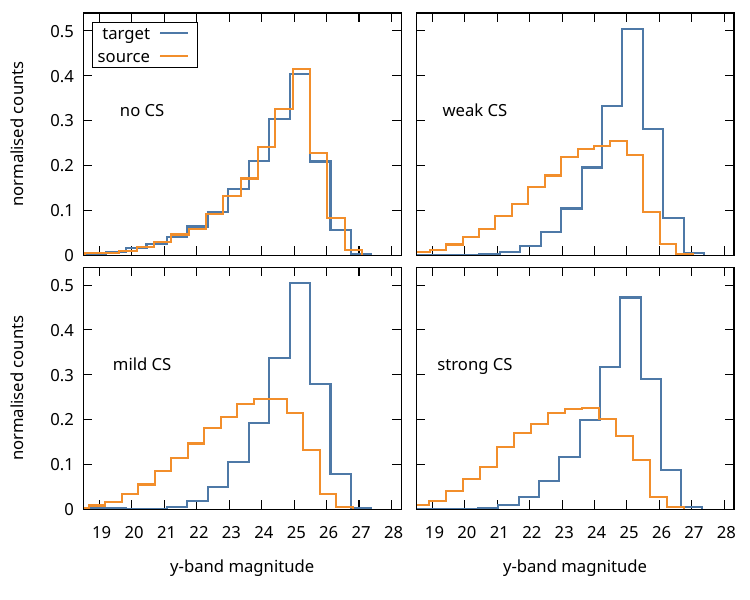}
\caption{Same as Fig.~\ref{fig:u-band} but for the $y$-band magnitudes.}
\label{fig:y-band}
\end{figure}

As mentioned in Sec.~\ref{sec:cs}, the introduction of covariate shift by rejection sampling on the $r$-band induces a shift also in the redshift distributions for source and target. Naturally, the effect is not limited to redshift, but also extends to other photometric bands. We plot here the distributions for all photometric bands (except for the $r$-band, already shown in Fig.~\ref{fig:test-train-histograms}), specifically Fig.~\ref{fig:u-band} for the $u$-band, Fig.~\ref{fig:g-band} for the $g$-band, Fig.~\ref{fig:i-band} for the $i$-band, Fig.~\ref{fig:z-band} for the $z$-band and Fig.~\ref{fig:y-band} for the $y$-band. As in the main text, orange histograms represent the source dataset, while we show in blue the target dataset.

\section{Stratification results}
\label{app:stratification-results}

\begin{figure}[t!]
\centering
\includegraphics[width=0.95\columnwidth]{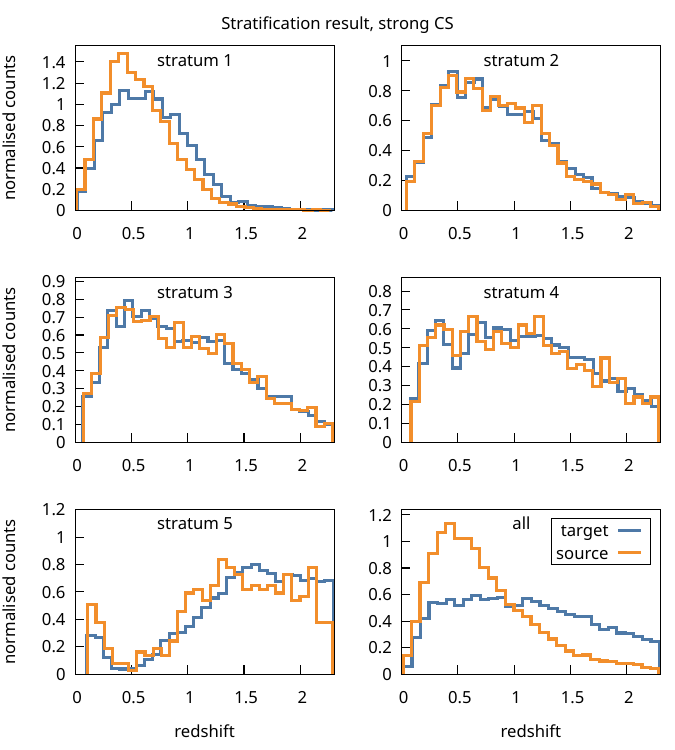}
\caption{Same as Fig.~\ref{fig:rband-strata} but for the redshift distributions. We note that this diagnostic is not possible in practice with real source/target data, as $z$ is not observed in the target dataset.}
  \label{fig:redshift-strata}
\end{figure}

We show here the effectiveness of the stratification based on propensity scores in removing covariate shift within each stratum for the redshift distributions. We focus on the worst case scenario, i.e.,  the strong covariate shift case. Each panel of Fig.~\ref{fig:redshift-strata} refers to one of the five strata, except for the bottom right panel which shows the redshift distributions for the full source and target sets. Blue histograms refer to the source set, while orange histograms refer to the target set.

\clearpage
\section{Toy model for PIT}
\label{app:pit-toy-model}

In this section, we illustrate a simple toy example to demonstrate the effect on the PIT histograms when overestimating (or underestimating) the uncertainties/errorbars of (conditional) density estimates.

\begin{figure*}
\centering
\includegraphics[width=0.33\textwidth]{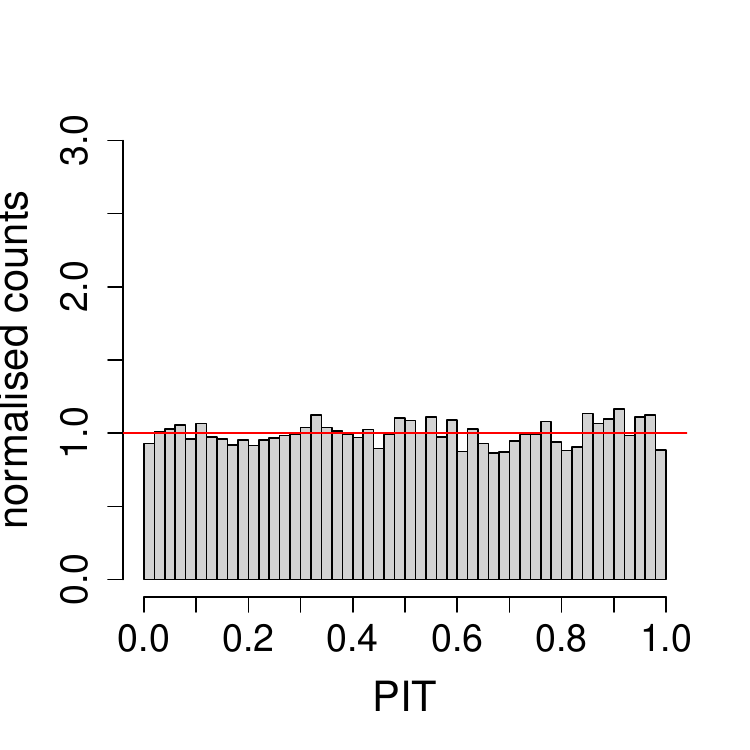}
\includegraphics[width=0.33\textwidth]{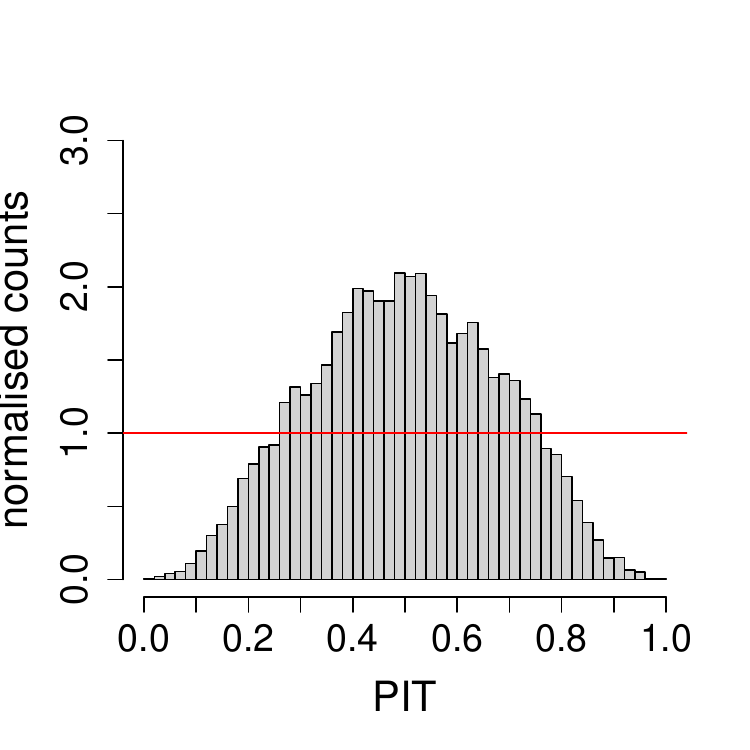}
\includegraphics[width=0.33\textwidth]{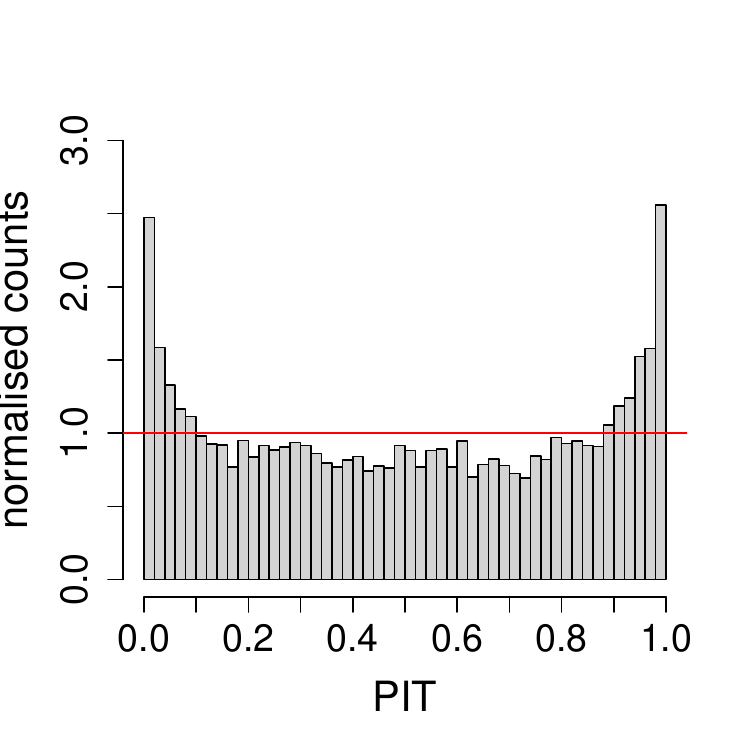}
\caption{PIT distributions illustrating the toy examples introduced in Appendix~\ref{app:pit-toy-model}.  Left panel: correct model; center panel: overestimated uncertainties; right panel: underestimated uncertainties.} \label{figure:PIT}
\end{figure*}

More precisely, we simulate data from  $X_i \overset{\text{i.i.d.}}{\sim} N(\mu = 0, \sigma = 1)$, with $i = 1,\dots,n$, obtaining $n=10.000$ independent samples. In a first step, we calculate the PIT assuming the correct model, i.e., $\text{PIT}_i = F_{X_i}(X_i)$, where $F_{X_i}(X_i)$ is the CDF of $X_i$ evaluated at $X_i$. As expected, the resulting histogram exhibits a standard uniform distribution, as illustrated in Fig.~\ref{figure:PIT} (left panel). 

In a next step, we keep the same data simulating process, however, we assume a model with much larger uncertainties (centered at the correct location). More precisely, we calculate the PITs assuming a much broader Gaussian distribution, i.e., we calculate $\text{PIT}_i = F_{X'_i}(X_i)$, where $F_{X'_i}(X_i)$ is the CDF of a random variable $X'_i \sim N(\mu = 0, \sigma =2)$, evaluated at $X_i$. 
Plotting the histogram of PIT values in Fig.~\ref{figure:PIT} (center panel), we find a large (symmetric) bump around 0.5, with less mass towards the edges zero and one.

Third, we assume a model that underestimates the uncertainties, i.e., we calculate the PITs via $\text{PIT}_i = F_{X''_i}(X_i)$, where $F_{X''_i}(X_i)$ is the CDF of a random variable $X''_i \sim N(\mu = 0, \sigma =0.8)$, evaluated at $X_i$. As illustrated in right panel of Fig.~\ref{figure:PIT}, this leads to large peaks at the edges (zero and one), indicating a large number of true values (from $X_i$) laying outside the assumed model distributions. (We note that similar peaks at the edges can be obtained by systematically shifting the center of the assumed model distribution away from the center of the underlying data-generating distributions).

We note that center panel of Fig.~\ref{figure:PIT} exhibits a similar pattern as the PIT histograms obtained for the \SLz conditional density estimates in Fig.~\ref{fig:pit}. The symmetric bump in the PIT histogram around 0.5 indicates that a large proportion of the conditional density estimates are well centered around the true values (as also suggested by the low RMSE and low mean error of \SLznospace, as presented in Table~\ref{tab:metrics}). However, there is an overestimation of uncertainties/errorbars in the conditional density estimates (overly conservative estimates).  On the contrary, underestimating uncertainties leads to large peaks of the PIT histograms at the edges (zero and one) indicating larger fractions of (catastrophic) outliers as it has been observed for the GPz estimates. Finally, while in the optimal case the PIT histograms exhibit a flat uniform distribution (as illustrated in left panel of Fig.~\ref{figure:PIT}), we argue that an overly conservative estimate of the uncertainties (with estimates well-centered around the true values) is more desirable than underestimating the uncertainties, which might lead to catastrophic outliers and a disproportionate number of estimates which do not cover the true values.

%%%%%%%%%%%%%%%%%%%%%%%%%%%%%%%%%%%%%%%%%%%%%%%%%%

% Don't change these lines
\end{document}